\documentclass[manuscript]{aastex63}
\usepackage{amsmath}
\usepackage{tabularx}
\usepackage{comment}
\usepackage{multirow}

\newcommand{\mathbfss}[1]{\textbf{\textsf{#1}}}
\newcommand{\mat}{\ensuremath{\mathbfss}}

\received{\today}
\submitjournal{ApJ}

\shorttitle{Whistler-regulated MHD}
\shortauthors{Drake et al.}

\begin{document}

\title{Whistler-regulated MHD: Transport equations for electron thermal conduction in the high $\beta$
  intracluster medium of galaxy clusters}

\correspondingauthor{James F. Drake}
\email{drake@umd.edu}

\author{J.~F.~Drake}
\affiliation{Department of Physics, University of Maryland,
  College Park, MD 20742, USA}
\affiliation{Institute for Physical Science and Technology,
  University of Maryland, College Park, MD 20742, USA}
\affiliation{Institute for Research in Electronics and Applied
  Physics, University of Maryland, College Park, MD 20742, USA}
\affiliation{Joint Space-Science Institute (JSI), College Park, MD 20742, USA}

\author{C.~Pfrommer}
\affiliation{Leibniz Institute for Astrophysics, Potsdam (AIP), An der Sternwarte 16, 14482 Potsdam, Germany}

\author{C.~S.~Reynolds}
\affiliation{Institute of Astronomy, University of Cambridge, Cambridge CB3 0HA, UK}

\author{M.~Ruszkowski}  
\affiliation{Department of Astronomy, University of Michigan, Ann Arbor, MI 48109, USA}

\author{M.~Swisdak}
\affiliation{Institute for Research in Electronics and Applied
  Physics, University of Maryland, College Park, MD 20742, USA}
\affiliation{Joint Space-Science Institute (JSI), College Park, MD 20742, USA}
 
  \author{A.~Einarsson}
  \affiliation{Department of Physics, University of Maryland,
  College Park, MD 20742, USA}
  
  \author{T.~Thomas}
\affiliation{Leibniz Institute for Astrophysics, Potsdam (AIP), An der Sternwarte 16, 14482 Potsdam, Germany}
\affiliation{Institute of Physics and Astronomy, University of Potsdam, Karl-Liebknecht-Str. 24/25, 14476 Golm, Germany}

\author{A.~B. Hassam}
\affiliation{Department of Physics, University of Maryland,
  College Park, MD 20742, USA}
\affiliation{Institute for Research in Electronics and Applied
  Physics, University of Maryland, College Park, MD 20742, USA}

\author{G.~T. Roberg-Clark}
\affiliation{Max-Planck-Institute for Plasmaphysics, 17491 Greifswald, Germany}



\begin{abstract}

Transport equations for electron thermal energy in the high $\beta_e$
intracluster medium (ICM) are developed that include scattering from
both classical collisions and self-generated whistler waves.  The
calculation employs an expansion of the kinetic electron equation
along the ambient magnetic field in the limit of strong scattering and
assumes whistler waves with low phase speeds
$V_w\sim{v}_{te}/\beta_e\ll{v}_{te}$ dominate the turbulent spectrum,
with $v_{te}$ the electron thermal speed and $\beta_e\gg1$ the ratio
of electron thermal to magnetic pressure. We find: (1)
temperature-gradient-driven whistlers dominate classical scattering
when $L_c>L/\beta_e$, with $L_c$ the classical electron mean-free-path
and $L$ the electron temperature scale length, and (2) in the whistler
dominated regime the electron thermal flux is controlled by both
advection at $V_w$ and a comparable diffusive term.  The findings
suggest whistlers limit electron heat flux over large regions of the
ICM, including locations unstable to isobaric condensation.
Consequences include: (1) the Field length decreases, extending the
domain of thermal instability to smaller lengthscales, (2) the heat
flux temperature dependence changes from $T_e^{7/2}/L$ to
$V_wnT_e\sim{T}_e^{1/2}$, (3) the magneto-thermal and heat-flux driven
buoyancy instabilities are impaired or completely inhibited, and (4)
sound waves in the ICM propagate greater distances, as inferred from obserations. This
description of thermal transport can be used in macroscale
ICM models.

\end{abstract}


\section{Introduction}\label{introduction}

Half of all baryons in the low-redshift Universe are in a hot phase, with temperatures in the range $T=10^{5-8}\,{\rm K}$ \citep{Macquart20}.  Much of this matter exists as very diffuse plasma in the filaments of the cosmic web, heated by photoionization from stars and active galactic nuclei \citep[AGN,][]{Mcquinn16} as well as electron-positron beam instabilities associated with TeV blazars \citep{Broderick12,Lamberts15}. Of more astrophysical significance are the hot baryons that have passed through cosmic accretion shocks \citep{Ryu03,Pfrommer06,Schaal16}, which caused the baryons to virialize in the dark matter halos of massive ($>L^*$) galaxies and galaxy clusters to form the hot circumgalactic medium (CGM) and hot intracluster medium (ICM), respectively.  

The ICM is particularly well studied since it has typical temperatures ($T\sim 10^7-10^8\,$K) and emission measures that make nearby galaxy clusters amongst the brightest objects in the extragalactic X-ray sky.  We now understand that the ICM dominates the baryon budget of clusters (containing 80\% of the baryons) and forms an approximately hydrostatic atmosphere in the dark matter potential.  In relaxed clusters (i.e., those not undergoing major mergers), the core regions of the ICM develop short cooling times and, left unchecked, a cooling catastrophe would occur resulting in significant star formation events ($100-1000\,{\rm M}_\odot$) and bright central galaxies (BCGs) with very massive ($10^{13}\,{\rm M}_\odot$) stellar components \citep{Sarazin1986,Fabian1994}.  The fact that BCGs are significantly less massive and never host such star forming events demonstrates that the ICM cooling must be balanced by a heat source \citep{Peterson2006}.  The current paradigm is that the jets from the central radio-loud active galactic nucleus (AGN) hosted by the BCG provide that heating \citep{Churazov00,Reynolds02}, although the precise mechanisms involved in thermalizing the AGN energy injection remain unclear \citep[see discussion in][]{Bambic19}. Thus, studies of galaxy clusters and the ICM give us a unique window on the AGN feedback processes that shape the most massive galaxies in the Universe. A key ingredient needed to disentangle these complex systems is the microphysics, and especially the transport properties, of the hot ICM.   

Thermal transport in the ICM is particularly important \citep{Binney81}.  Thermal conduction into the cooling ICM core from the hotter outer regions can reduce, but not eliminate, the need for AGN heating \citep{Stewart84,Bregman88,Zakamska03,Voigt04}.  Thermal transport may be crucial in dissipating weak shocks and acoustic modes driven by the AGN \citep{Fabian05,Zweibel2018} thereby thermalizing AGN injected energy.  Thermal transport is also important for modifying the local thermal instabilities responsible for the condensation of cold gas that, ultimately, fuels the AGN \citep{Yang16}.  But despite its importance, the basic physics of thermal transport in ICM-like plasmas has yet to be fully understood.  At typical densities ($n_e\sim 10^{-3}-10^{-1}\,{\rm cm}^{-3}$) and temperatures, the electron mean free path to ion scattering is $L_c\sim 0.1-1\,{\rm kpc}$ and thus only 1--3 orders of magnitude smaller than global scales. The ICM is magnetized with typical fields $B\sim 1-10\mu{\rm G}$, giving rise to electron gyro-radii $\rho_e\sim 0.1-1\,{\rm npc}$ many orders of magnitude smaller  than $\lambda$, implying that the transport will be highly suppressed in the cross-field direction and thus anisotropic.  Most current treatments of the ICM adopt a fluid description and take the thermal transport parallel to the local magnetic field to be described by the canonical theory of \cite{Spitzer56}.  However, the Spitzer ansatz assumes a high degree of electron-ion collisionality which is not obviously true for the ICM.  The fact that the ratio of thermal-to-magnetic pressure is large ($\beta\sim 100$) further raises the spectre of kinetic instabilities driven by pressure anisotropies and heat fluxes \citep{Gary00,Schekochihin05,Kunz14,Rincon15,Riquelme16,Komarov16}.

More recently the focus has shifted to the role of whistler waves that
are directly driven by electron heat flux in suppressing transport
\citep{Levinson92,Pistinner98,Roberg-Clark16,Roberg-Clark18,Komarov18}. Transport
suppression is dominated by oblique waves that can resonate with
whistlers propagating in the direction of the electron heat flux through
the Landau ($n=0$) and ``anomalous'' ($n=-1, -2, ...$) resonances $\omega -k_\parallel
v_\parallel -n\Omega_e=0$, where
$\Omega_e=eB/m_ec$ is the electron cyclotron
frequency. Particle-in-cell (PIC) simulations carried out in two dimensions
revealed that the whistler fluctuations reach large amplitude and are
dominated by wavenumbers with $k\rho_e\sim 1$, where $\rho_e=v_{te}/\Omega_e$ is the
electron Larmor radius. The dispersion relation for whistlers with
wavelength $kd_e\sim d_e/\rho_e\ll 1$ is given by
\begin{equation}
  \omega=k_\parallel kd_e^2\Omega_{e}\ll \Omega_{e}
  \label{whistler}
\end{equation}
with $d_e$ the collisionless skin depth. The parallel phase speed
$V_w=\omega/k_\parallel$ of whistlers in this regime therefore scales
like $v_{te}/\beta_e\ll v_{te}$, where $\beta_e=8\pi nT_e/B^2$ is the
ratio of electron thermal to magnetic field pressure. Note that in the
definition of $\beta_e$, and in the remainder of this paper, the electron
temperature will be expressed in energy units. The simulations
revealed that electron scattering was strong enough that hot
electrons were constrained to move with the whistler phase speed down
the temperature gradient, reducing the heat flux well below the
free-streaming value $Q_0\sim nT_ev_{te}$ to a level that scaled as
$Q_0/\beta_e$ \citep{Pistinner98,Roberg-Clark18,Komarov18}. Interestingly, the
inverse scaling of the electron heat flux with $\beta_e$ has been documented
in the solar wind \citep{Tong18,Tong19}. Moreover, recent observations
from the Parker Solar Probe, revealed large amplitude, oblique
whistler waves embedded in regions of depressed magnetic field
intensity where the local $\beta_e$ was high \citep{Agapitov20,Cattell21}. The waves were associated with the $n=-1$ anomalous resonance and scattered the field-aligned strahl electrons into the halo population with a broad range of pitch angles \citep{Cattell21a}.

Due to computational constraints, PIC simulations exploring
heat flux suppression employ ambient temperature gradients
that are artificially large, with gradient scale lengths that
are only hundreds of electron Larmor radii. Thus, while the simulations
suggested that the scaling of the suppressed heat flux was insensitive
to the ambient gradient, the extrapolation to physical systems
requires that the results scale as expected over many orders of
magnitude. Further, simulations to date have been carried out in the
limit where classical collisions are negligible so that their impact on the onset of the whistler instability is uncertain. Thus, the transition from whistler-limited to classical-collision-limited transport has not been explored beyond the construction of
{\it ad hoc} connection formulas to bridge the two limits \citep{Komarov18}.

The goal of the present work is to develop a set of transport
equations that treat macroscale systems while simultaneously describing
the transition from classical- to whistler-limited transport. Two characteristics of scattering by
whistler waves make the calculation possible. First, whistlers scatter
electrons on constant energy surfaces in the frame moving with the
whistler phase speed $V_w$ along the ambient magnetic field. Hence, the
whistler scattering can be formulated as a pitch angle scattering
operator moving at $V_w$. Second, because
$V_w=v_{te}/\beta_e\ll v_{te}$, the kinetic equation for electrons can
be solved by an ordering in which $V_w/v_{te}\ll 1$. At the same time,
the total scattering rate, $\nu (v)=\nu_w(v)+\nu_{ei}(v)$ includes both classical and whistler scattering and is ordered so that
the mean-free-path $L_\nu=v_{te}/\nu\sim L/\beta_e$ is short compared with
the ambient temperature scale length along the magnetic
field, $L$. To simplify the calculation, the classical collisions are also
treated by a simple pitch-angle scattering operator, which is, of
course, valid for electron-ion but not for electron-electron
collisions. The final transport equations for the electron thermal
energy result from solving the kinetic equation for the electron
distribution $f(x, v, \zeta, t)$ to second order in the small parameter
$\epsilon \sim V_w/v_{te} \sim L_\nu/L$ with no assumptions made about
the relative sizes of the classical and whistler scattering
rates. Here, $\zeta=v_\parallel/v$ is the cosine of the pitch angle.

The calculation that follows is not without assumptions. Specifically we  assume that once whistlers begin to grow 
they are able to scatter electrons through the full range of pitch
angle. For oblique whistlers this is reasonable because of the multiple anomalous resonances and if the spectrum of waves is sufficiently
broad. The assumption that the rate of scattering is independent on the pitch angle is based on the idea that electron pileup at a specific pitch angle would produce instabilities that resonate with electrons at that pitch angle and therefore facilitate scattering. Thus, the assumption is not that the whistler amplitudes are large, which is not expected in a system with very large ambient gradient scale lengths, but that resonant overlap in the quasilinear sense is a consequence of the broad wave spectrum -- there are enough waves so that all electrons can undergo resonant interactions. We emphasize also that in the solar wind, where gradients scales are also large, observations reveal that the strahl electrons are scattered over a broad range of pitch angles \citep{Cattell21a}. Finally, in the present calculation we assume that heat flux driven whistlers dominate electron scattering. Other scattering mechanisms associated with, for example, pressure anisotropy driven mirror modes have also been proposed \citep{Komarov16} but are not included in the present calculation. 

The ambient ion velocity $U_i$ is included in the equations for
completeness with the ordering $\epsilon \sim U_i/v_{te}$. Transport
is described by coupled equations for the electron density $n(x,t)$,
the electron pressure $P_e(x,t)$, and the energy density of propagating whistler waves such that the total energy is
conserved. The calculation parallels that carried out by Braginskii
for the case of classical collisions \citep{Braginskii65,Hassam80} with
functional characteristics that parallel the equations
describing cosmic ray transport limited by Alfv\'en wave scattering
\citep{Kulsrud69,Zweibel13,Thomas19}.

A surprise is that the final equations describe both the drive of the
whistlers by the release of energy associated with particle scattering
by the waves and the saturation of the instability when
the scattering rate exceeds a threshold. The onset of whistler growth
is controlled by classical collisions and requires $L_c>L/\beta_e$
with $L_c=v_{te}/\nu_{ei}(v_{te})$ the classical mean-free-path. The
saturation of the instability takes place when $L_w\sim L/\beta_e$
with $L_w\sim v_{te}/\nu_w(v_{te})$ the whistler scattering
mean free path. Thus, the scattering mean free path of electrons in a
high $\beta_e$ medium is always much smaller than the ambient scale
length of the electron temperature even in the nominally collisionless
domain. A second surprise is that the electron heat flux in the regime
in which whistler scattering dominates classical collisions is not
simply given by advection at the whistler phase speed $V_w$  but is controlled by a combination of
advection and diffusion, which are of the same order. This follows
from the diffusive heat flux, which is of order
\begin{equation}
  n\frac{v_{te}^2}{\nu_w(v_{te})}\frac{T_e}{L}\sim nT_ev_{te}\frac{L_w}{L}\sim nT_e\frac{v_{te}}{\beta_e}\sim nT_eV_w.
  \label{heatfluxscaling}
  \end{equation}
That the electron heat flux has contributions from both advection and
diffusion was overlooked in earlier papers describing the results
of simulations \citep{Roberg-Clark18,Komarov18}, but is consistent with
the analogous equations for the transport of cosmic rays
\citep{Thomas19}.

The manuscript is organized as follows: in Sec.~\ref{overview} we
present an overview of the transport equations, including a
description of their basic properties; in Sec.~\ref{derivation} we
present the derivation of the transport equations with a discussion of
the basic assumptions of the model; in Sec.~\ref{computations} we
present the results from computations of the propagation of an electron
heat pulse that contrasts the results of transport in a classical
collisional dominated regime with one with whistler scattering;
and in Sec.~\ref{conclusions} we discuss the implications of our
results for the ICM. Finally in the Appendix we present a set of equations that combines the new description of electron thermal transport with the conventional magnetohydrodynamic (MHD) equations. The resulting Whistler-regulated MHD equations (W-MHD) are suitable for describing the full dynamics of ICM plasma over the full range of classical collisionality with none of the conventional constraints on mean free path.   

\section{Overview of the Electron Transport Equations}\label{overview}
Here we present an overview of the equations describing electron
transport. The scattering operator representing classical collisions
is energy dependent but only scatters the electrons in pitch angle and
therefore only accurately models electron-ion collisions. The whistler
scattering operator is also energy-dependent and scatters in pitch
angle in a frame moving with a single velocity $V_w\ll v_{te}$. The
energy dependence of the whistler scattering rate is essential to
correctly describe a finite heat flux while at the same time
maintaining zero net current. Below we write three equations that
describe the plasma density, the
electron energy, and the energy associated with electron heat-flux
driven whistler waves. The continuity equation is given by
\begin{equation}
  \frac{\partial}{\partial t}n+\nabla_\parallel nU=0
  \label{continuity}
\end{equation}
with $U$ the ion drift speed along the ambient large-scale magnetic
field (not including the whistler magnetic fluctuations). Because of
the zero net current condition $U$ is also the mean electron drift
speed.  The equation
for the electron energy is
\begin{equation}
  \frac{\partial}{\partial t}\left(\frac{3}{2}nT\right)+\nabla_\parallel
  Q-U\nabla_\parallel nT=V_w\left(\alpha_{10}\frac{\nu_{we}}{\nu_e}n\nabla_\parallel T+F_w\right)
  \label{energy_electrons}
  \end{equation}
  where the parallel heat flux $Q$ is
  \begin{equation}
    Q=\left( \frac{5}{2}U+\alpha_{10}\frac{\nu_{we}}{\nu_e}V_w\right) nT-\kappa_e\nabla_\parallel T.
    \label{heatflux}
  \end{equation}
It includes the traditional advection of the enthalpy, $5nT/2$, with the
fluid velocity $U$ and contributions from advection by the whistler waves
(proportional to $V_w$) as well as from thermal 
conduction with conductivity $\kappa_e$,
  \begin{equation}
    \kappa_e=\alpha_{12}\frac{nT}{m_e\nu_e}.
      \label{conductivity}
  \end{equation}
In Eqs.~(\ref{energy_electrons})-(\ref{conductivity}) the $\alpha_i$ parameters
labeled with subscripts are dimensionless and of order
unity. They arise from averages of combinations of the scattering rates
and powers of the particle velocity $v$ over Maxwellian
distributions. Explicit expressions are given in Table
\ref{alpha_params}.   The total scattering rate $\nu_e=\nu_{ei}^e+\nu_{we}$ is evaluated at the electron thermal speed $v_{te}$ and
  \begin{equation}
    \nu_{ei}^e=\frac{4\pi e^4n\Lambda}{m_e^2v_{te}^3} \qquad \text{and} \qquad \nu_{we}=0.1\Omega_{e}\frac{\varepsilon_w}{\varepsilon_B}. 
      \label{nu}
  \end{equation}
The whistler scattering rate $\nu_{we}$ is given by the quasilinear
form with $\varepsilon_w$ and $\varepsilon_B$ the energy densities of
the whistler waves and the large scale magnetic field $B$,
respectively \citep{Lee82,Schlickeiser89}, and $\Lambda$ is the Coulomb logarithm. We note that because we have discarded
electron-electron the parallel conduction in Eq.~(\ref{conductivity})
does not reduce to the Spitzer value \citep{Hassam80}. As discussed in
the Appendix, the parameter $\alpha_{12}$ can be corrected to produce
the Spitzer value. In this overview of the transport equations we assume that the temperature gradient is everywhere negative so that only whistlers traveling in the positive direction along ${\bf B}_0$ are unstable. The general equations presented in the Appendix include waves propagating in both directions along ${\bf B}_0$. In the absence of scattering by whistlers the
equation for $nT$ reduces to the one-dimensional (1D) Braginskii equation
\citep{Braginskii65}. 

The advection of electron energy associated with the whistler wave
propagation has been documented in PIC simulations
\citep{Roberg-Clark18,Komarov18}. However, the whistler-driven
advective heat flux becomes small when classical collisions
dominate those associated with whistlers. This behavior was not
explored in PIC simulations since classical collisions were not
included. The thermal conduction term takes the classical form, being
inversely proportional to the collision rate $\nu_e$. The collision
rate includes both classical and whistler-driven scattering so that,
even in the absence of classical collisions, whistler waves also
produce diffusive transport. This result is consistent with the
measured parallel diffusion of electrons scattered by heat-flux-driven
whistlers in simulations \citep{Roberg-Clark18,Komarov18}. However, the diffusive
contributions to heat flux associated with whistler scattering were
missed in previous discussions of these simulations. A wave-driven flux with a form
similar to that of Eq.~(\ref{heatflux}) also appears in  equations
describing the transport of cosmic rays
\citep{Kulsrud69,Zweibel13,Thomas19}.

The two terms on the right side of Eq.~(\ref{energy_electrons}), which
are proportional to $V_w$, describe the extraction of electron thermal
energy by whistler waves and the heating of electrons associated with
whistler scattering. The energy extraction term is proportional to the
local temperature gradient and is negative for waves propagating down
the gradient, corresponding to energy extraction from the gradient by
whistlers. The dimensionless coefficient $\alpha_{10}$ leading the
extraction term is only nonzero when the whistler scattering rate
$\nu_w$ depends on the electron
energy. For a scattering rate that has the powerlaw dependence,
$\nu_w\propto v^\gamma$ with $\gamma > 0$, $\alpha_{10}$ is
proportional to $\gamma$ in the limit of both strong and weak
classical scattering. The powerlaw form for the energy dependence of
$\nu_w$ results from a quasilinear model in which electrons with
higher energies resonate with longer wavelength modes (see
Eq.~\ref{whistler}) and longer wavelength modes have larger
amplitudes in a system undergoing an energy cascade
\citep{Lee82,Schlickeiser89,Schlickeiser98}. Further discussion of the
appropriate value for $\gamma$ is presented in Sec.~\ref{derivation}.
The term proportional to $F_w$ in Eq.~(\ref{energy_electrons}) arises
from the drag force between the whistlers and electrons and is given by
\begin{equation}
  F_w=\alpha_{11}m_en\nu_{we}V_w
  \label{equation:Fw}
\end{equation}
It is positive and drives electron heating.  It is a consequence of the linear resonant damping of whistlers, which heats the electrons. This is discussed further in the context of the equation for the growth of whistlers by the ambient heat flux below. The balance between energy extraction
from the temperature gradient and from the damping of whistlers
by electrons that describes both the onset of whistler wave growth in a
medium with classical collisions and the saturation of the whistler
wave energy. 

An evolution equation for the whistler wave energy $\varepsilon_w$ is
required to close Eq.~(\ref{energy_electrons}). The equation for
$\varepsilon_w$ includes advection both by the plasma fluid with
velocity $U$ and by the whistlers with velocity $V_w \sim
v_{te}/\beta_e$ as well as drive and loss contributions associated with the
extraction and heating terms on the right side of
Eq.~(\ref{energy_electrons}).  Balancing the energy gain of the whistlers with the
energy loss from electrons yields the evolution equation
\begin{equation}
  \frac{\partial }{\partial t}\varepsilon_w + 2\nabla_\parallel (V_w+U)\varepsilon_w-U\nabla_\parallel\varepsilon_w=
  -V_w\left(\alpha_{10}\frac{\nu_{we}}{\nu_e}n\nabla_\parallel T+F_w\right),
  \label{energy_whistlers}
\end{equation}
where the second term on the left side of the equation is the Poynting
flux and the third term is associated with the whistler
pressure acting on ions. There is a corresponding term in the ion kinetic energy equation
(see Eq.~(\ref{eps_kin})) so that the ions and whistlers can exchange
energy through this pressure. Thus, the sum of
Eqs.~(\ref{energy_electrons}) and (\ref{energy_whistlers}) plus the
equation for ion kinetic energy yields an equation for overall energy
conservation.

An important scientific goal is to establish the conditions under
which whistler waves in a high $\beta_e$ system begin to grow even
when there are ambient classical collisions. The onset condition for
whistler amplification can be obtained from
Eq.~(\ref{energy_whistlers}) by taking the limit in which the whistler
wave energy $\varepsilon_w$ is small so that the whistler scattering
rate $\nu_{we}$ in Eq.~(\ref{nu}) is much smaller than $\nu_{ei}$. The
resulting equation for the rate of growth $\gamma_w$ of the wave energy is given by
\begin{equation}
  \gamma_w = -nV_w\frac{\partial\nu_{we}}{\partial\varepsilon_w}\left(\alpha_{10}\frac{\nabla_\parallel T}{\nu_{ei}^e}+\alpha_{11}m_eV_w\right).
  \label{growthrate}
\end{equation}
The drive term proportional to $\nabla_\parallel T$ extracts energy from electrons (see Eq.~(\ref{energy_electrons})) to destabilize the whistlers. The damping term can be evaluated explicitly using Eq.~(\ref{nu}) to evaluate $\nu_{we}$. The result for the whistler damping $\gamma_{wd}$
\begin{equation}
   \frac{\gamma_{wd}}{\omega_0} \sim 0.1 \frac{V_w}{v_{te}}\beta_e,
\end{equation}
where $\omega_{0}=\Omega_e/\beta_e$ is the characteristic whistler wave frequency. This scaling for whistler wave damping is identical to that obtained from the resonant interaction of electrons with oblique whistlers from kinetic theory. The onset condition for whistler growth is insensitive to the details of the scattering rate and is given by
\begin{equation}
  - \nabla_\parallel T>\frac{\alpha_{11}}{\alpha_{10}}m_e\nu_{ei}^eV_w
  \label{growthrate_threshold}
\end{equation}
or $\beta_eL_c/L >1$. A similar result came from a quasilinear analysis of whistler stability \citep{Pistinner98}. Above threshold the growth rate $\gamma_w$ of the whistler wave energy scales as
\begin{equation}
  \gamma_w \sim 0.1 \Omega_{e}\frac{L_c}{L}.
  \label{growthrate_scaling}
\end{equation}
A measure of the strength of whistler growth is given by a comparison
with the rate at which whistlers transit down the temperature gradient
$V_w/L$. The whistler growth rate exceeds the transit rate for
$\beta>10\rho_e/L$, which is easily satisfied since $\rho_e$ is many orders
of magnitude smaller than the macroscale $L$. The implication is that
once the threshold for whistler growth is exceeded, the whistlers will
rapidly reach finite amplitude. 

The saturation of the whistler waves is controlled by the balance
between the drive and damping terms on the right side of
Eq.~(\ref{energy_whistlers}). The drive term becomes independent of
$\nu_{we}$ when $\nu_{we}>\nu_{ei}$ while the dissipation term, which
is proportional to $F_w$ and $\nu_{we}$ continues to increase as
whistler growth continues. Thus, whistler growth continues until the
drive and dissipation terms on the right side of
Eq.~(\ref{energy_whistlers}) balance. This balance gives the rate of
whistler scattering at saturation,
\begin{equation}
  \nu_{we} = -\frac{\alpha_{10}}{\alpha_{11}} \frac{1}{m_eV_w} \nabla_\parallel T,
  \label{nu_whistler}
\end{equation}
which yields the scaling for the whistler scattering rate
$\nu_{we}\sim \beta_ev_{te}/L$ and the whistler scattering
mean-free-path $L_w=v_{te}/\nu_{we}\sim L/\beta_e$. Thus, even when
classical electron-ion collisions are weak, whistler driven scattering
reduces the electron mean-free-path to a scale length that is small
compared with the ambient temperature scale length $L$.  Further, the saturated value of $\nu_{we}$ is insensitive to the details of the whistler growth rate in Eq.~(\ref{growthrate}). At saturation the whistler wave energy density $\varepsilon_w$ can be
calculated by equating the expressions for $\nu_{we}$ in
Eqs.~(\ref{nu}) and (\ref{nu_whistler}), which yields $\varepsilon_w
\sim 10(\rho_e/L)nT\ll nT$. Thus, the whistler energy density remains
small compared with the electron thermal energy density \citep{Komarov18}.

The fast growth rate of whistlers compared with the transit time of
the waves across the system suggests that the whistlers may reach an
equilibrium even when the wave drive only weakly exceeds the threshold
for instability. This state can be explored by discarding the time
derivative in Eq.~(\ref{energy_whistlers}) so that the balance between
the drive and dissipation terms on the right side of the equation must
scale as $\nabla_\parallel V_w\varepsilon_w\ll V_wnT/L$. Therefore,
in steady state the drive and dissipation terms on the right side of
Eq.~(\ref{energy_electrons}) must balance and their difference must be
negligible compared with the contribution from
the heat flux since $\nabla_\parallel Q\sim V_wnT/L$. The heat flux
$Q$ therefore dominates the dynamics of the electron energy
equation and $\nu_e$ is given by the balance of the drive and
dissipation terms,
\begin{equation}
  \nu_e=\nu_{ei}^e+\nu_{we}=-\frac{\alpha_{10}}{\alpha_{11}}\frac{1}{m_eV_w}\nabla_\parallel T.
    \label{nu_total}
\end{equation}
This expression for $\nu_e$ is valid as long as the onset condition
for whistler growth in Eq.~(\ref{growthrate_threshold}) is satisfied so that
$\nu_{we}$ is positive. This expression for $\nu_e$ can then be inserted into the expression for the heat flux in Eq.~(\ref{heatflux}), which yields
\begin{equation}
  Q=\left[\frac{5}{2}U+\left(\alpha_{10}\left(1-\frac{\nu_{ei}^e}{\nu_e}\right)+\frac{\alpha_{11}\alpha_{12}}{\alpha_{10}}\right)V_w\right]nT.
  \label{heatflux_total}
\end{equation}
This expression is valid above the threshold for whistler growth and
reveals that, in the regime dominated by whistlers, the proportionality
of the scattering rate to $\nabla_\parallel T$ causes the expected
diffusive driven energy flux to become advective. At marginal
stability where $\nu_e=\nu_{ei}$ the first term in the heat flux in
Eq.~(\ref{heatflux_total}) is zero and the second term in the heat
flux reduces to the classical diffusive form. It remains to be seen
whether the form of the heat flux given in Eq.~(\ref{heatflux_total})
is useful in a system where the whistler transitions from a stable to
unstable domain, which will happen in any system where the electron
temperature peaks so that $\nabla_\parallel T$ is zero in a local
region and there is no whistler drive.

\section{Derivation of the electron transport equations}\label{derivation}
Having presented an overview of the electron transport equations, we
proceed to outline the assumptions, ordering and derivation. The focus here is on the electron distribution function and resulting moments and the wave equation for the whistler waves. Because the ion momentum and energy equations are largely unchanged from the traditional MHD description (barring a small force term associated with the whistler pressure acting on the ions) the coupling to the ion momentum and energy equations are deferred to Appendix~\ref{appendix} where the complete set of MHD equations with coupled whistler dynamics is presented. 
\subsection{Chapman-Enskog solution of the electron kinetic equation}
We begin with an equation for the distribution
function $f(x, v_\parallel, v_\perp, t)$, where $x$ is the space
variable parallel to the ambient magnetic field ${\bf B}_0$ and
$v_\parallel$ and $v_\perp$ are the velocities parallel and
perpendicular to ${\bf B}_0$. Since the whistler waves in a high $\beta_e$
system have $k\rho_e \sim 1$ so that the whistler wave frequency is
well below the electron cyclotron frequency (see
Eq.~\ref{whistler}), the electrons remain gyrotropic. The kinetic equation for electrons becomes
\begin{equation}
  \frac{\partial}{\partial t}f+v_\parallel\nabla_\parallel f-\frac{e}{m_e}E_\parallel\frac{\partial }{\partial v_\parallel}f=C(f),
  \label{equation:f}
\end{equation}
where a parallel electric field $E_\parallel$ has been included to maintain zero net current. The zero current condition is valid even in a system with complex three-dimensional magnetic fields as long as the kinetic scales are ordered out of the dynamics \citep{Drake19,Arnold19}. The collision operator includes scattering by both classical electron-ion collisions and whistler waves
\begin{equation}
  C(f)=C_{ei}(f)+C_w(f)
  \label{collision_total}
\end{equation}
with
\begin{equation}
  C_{ei}=\frac{1}{2}\nu_{ei}(|{\bf v}-{\bf U}|)\frac{\partial}{\partial {\bf v}}\cdot\left( \vec{\vec{I}}|{\bf v}-{\bf U}|^2-({\bf v}-{\bf U})({\bf v}-{\bf U})\right)\cdot\frac{\partial }{\partial {\bf v}}
  \label{collision_ei}
\end{equation}
and    
\begin{equation}
  C_w=\frac{1}{2}\nu_w(|{\bf v}-{\bf U}-{\bf V}_w|)\frac{\partial}{\partial {\bf v}}\cdot\left( \vec{\vec{I}}|{\bf v}-{\bf U}-{\bf V}_w|^2-({\bf v}-{\bf U}-{\bf V}_w)({\bf v}-{\bf U}-{\bf V}_w)\right)\cdot\frac{\partial }{\partial {\bf v}}
  \label{collision_w}
\end{equation}
describing the respective types of scattering. To simplify the derivation, we include here only whistlers propagating in the positive direction, which is valid for a locally negative temperature gradient. The generalization to counter-propagating whistlers is straightforward and the results are presented in the Appendix. The collision operators in Eqs.~(\ref{collision_ei}) and (\ref{collision_w}) describe scattering only in
pitch angle -- in the frame moving with the net parallel plasma drift $U$
in the case of $C_{ei}$ and with respect to the whistler wave frame
$U+V_w$ in the case of whistler waves. Both scattering rates have
energy dependencies with $\nu_{ei}\propto v^{-3}$ and $\nu_w\propto
v^\gamma$. It will follow from the solutions to Eq.~(\ref{equation:f})
that the energy dependence of $\nu_w$ is crucial for maintaining a net
whistler driven advective heat flux while at the same time
maintaining zero net current. An important assumption in the model for
whistler scattering is that the scattering preserves energy in the
whistler wave frame, which follows from the fact that, for long wavelength oscillations in that frame, the electric field vanishes. An
additional assumption, however, is that $\nu_w$ is independent of
pitch angle. Such a model is supported by PIC simulations of
heat-flux-driven whistlers in which the particles were isotropized in
pitch angle. On the other hand, the amplitudes of whistlers in a real
system are likely to saturate at much lower values than in the
simulations because the temperature gradient scale lengths in real
systems are far larger. As
discussed in Sec.~\ref{overview}, the saturation amplitude of the
whistler fluctuations $\varepsilon_w$ scales as $nT\rho_e/L\ll
nT$. Quasilinear models of electron scattering based on an assumed
spectrum of waves yield scattering rates that depend on the pitch
angle through $\zeta=v_\parallel/v$. However, we argue that the spectrum of
heat-flux-driven waves will evolve so that the scattering rate is
insensitive to $\zeta$ since any pileup of the distribution function
with pitch angle would lead to local growth of whistlers which would
enhance the local scattering. For this reason, the whistler scattering
rate is taken to be independent of pitch angle.

Equation (\ref{equation:f}) is solved to second order in the parameter
$\epsilon$ discussed at the end of
Sec.~\ref{introduction}. Fundamental to the ordering is the assumption
that the collisional mean free path is short compared with the parallel
system scale length $L$ and that both the mean drift speed $U$ and the
whistler phase speed $V_w$ are small compared with the thermal speed
so that the collision operators as well as $f$ can be expanded in
powers of $\epsilon \sim V_w/v_{te} \sim U/v_{te}$. Thus, we begin by writing Eq.~(\ref{equation:f}) to lowest order,
\begin{equation}
  0=C_0(f_0),
  \label{equation:f0}
\end{equation}
with
\begin{equation}
  C_0=\frac{1}{2}(\nu_{ei}+\nu_w)\frac{\partial}{\partial {\bf v}}\cdot\left( \vec{\vec{I}}v^2-{\bf v}{\bf v}\right)\cdot\frac{\partial }{\partial {\bf v}},
  \label{C0}
\end{equation}
whose solution is a Maxwellian with zero drift,
\begin{equation}
  f_0=n_0\left(\frac{m_e}{2\pi T_0}\right)^{3/2}e^{-m_ev^2/2T_0}.
  \label{f0}
\end{equation}
Technically, $C_0$ only scatters in pitch angle so $f_0$ could be
written as any function of $v^2$. However, we take $f_0$ to be a
Maxwellian. To first order, the derivatives on the left side of Eq.~(\ref{equation:f}) need to be included,
\begin{equation}
  \frac{\partial}{\partial t}f_0+v_\parallel\nabla_\parallel f_0-\frac{e}{m_e}E_\parallel\frac{\partial }{\partial v_\parallel}f_0=C_0(f_1)+C_1(f_0),
  \label{equation:f1}
\end{equation}
where
%
\begin{equation}
  \begin{aligned}
  C_1= & \frac{1}{2}\nu_w(v)\frac{\partial}{\partial {\bf v}}\cdot\left( -2\vec{\vec{I}}{\bf         v}\cdot ({\bf U}+{\bf V}_w)+{\bf v}({\bf U}+{\bf V}_w)+({\bf U}+{\bf V}_w){\bf              v}\right)\cdot\frac{\partial }{\partial {\bf v}} \\ 
       &  -\frac{\partial\nu_w}{\partial v^2}{\bf v}\cdot ({\bf U}+{\bf V}_w)\frac{\partial      }{\partial {\bf v}}\cdot (\vec{\vec{I}}v^2-{\bf v}{\bf v})\cdot\frac{\partial          }{\partial {\bf v}} \\ 
        & +(\nu_w\rightarrow\nu_{ei}, {\bf V}_w\rightarrow 0)
  \label{C1}
  \end{aligned}
\end{equation}
Both $C_0$ and $C_1$ preserve the number density and the kinetic energy when averaged over velocity and the mean drift of $f_0$ is zero so
\begin{equation}
  \frac{\partial n_0}{\partial t}=0, \;\;\; \frac{3}{2}\frac{\partial P_0}{\partial t}=0
  \label{equation:n0,T0}
\end{equation}
with $P_0=n_0T_0$. Thus, the time derivative of $f_0$ in Eq.~(\ref{equation:f1}) can be neglected. The various derivatives acting on $f_0$ (from $\nabla_\parallel$, $\partial /\partial {\bf v}$ and $C_0$) can be readily evaluated. The inversion of $C_0(f_1)$ is simplified by noting that $f_1=f_1(x, v, \zeta, t)$ so that
\begin{equation}
  C_0(f_1)=\frac{\nu }{2}\frac{\partial }{\partial \zeta}(1-\zeta^2)\frac{\partial }{\partial \zeta } f_1,
  \label{equation:C0(f1)}
\end{equation}
which is then readily inverted to yield the solution for $f_1$,
\begin{equation}
  f_1=-\frac{1}{\nu}\frac{v\zeta f_0}{T_0}\left[ \frac{1}{n_0}\nabla_\parallel P_0
  +\left( \frac{m_ev^2}{2T_0}-\frac{5}{2}\right)\nabla_\parallel T_0+eE_\parallel-m_e\left( \nu_w(V_w+U)+\nu_{ei}U\right) \right].
  \label{f1}
  \end{equation}
The first order electron drift velocity $U_1$ can then be evaluated by
multiplying $f_1$ by $v_\parallel =v\zeta$ and integrating over
velocity. The result is Ohm's law,
\begin{equation}
  0=m_e\nu_e(U_1-U)=-\alpha_1\left(\frac{1}{n_0}\nabla_\parallel P_0+eE_\parallel\right)-\alpha_2\nabla_\parallel T_0+\alpha_3m_e\nu_{we}V_w,
  \label{equation:ohm}
\end{equation}
where we have invoked the zero current condition to require
$U_1=U$. This equation then determines $E_\parallel$. The
dimensionless parameters $\alpha_1$, $\alpha_2$, and $\alpha_3$ are given in Table
\ref{alpha_params}. The collision rates in Eq.~(\ref{equation:ohm})
are evaluated at the electron thermal speed,
$\nu_e=\nu_{we}+\nu_{ei}^e$ with $\nu_{we}=\nu_w(v_{te})$ and
$\nu_{ei}^e=\nu_{ei}(v_{te})$. The term proportional to
$\nabla_\parallel T_0$ in Eq.~(\ref{equation:ohm}) is the thermal
force and is only nonzero when the velocity dependence of the
scattering rate is included. The first order electron heat flux $Q_1$ can
also be evaluated by multiplying $f_1$ by $m_ev^2v_\parallel /2$ and
averaging over velocity. After eliminating $E_\parallel$ using Eq.~(\ref{equation:ohm}), the result for $Q_1$ is given by
\begin{equation}
  Q_1=\left( \frac{5}{2}U+\alpha_{10}\frac{\nu_{we}}{\nu_e}V_w\right) n_0T_0-\kappa\nabla_\parallel T, 
  \label{Q1}
\end{equation}
with $\kappa$ given in Eq.~(\ref{conductivity}). When the whistler
scattering rate is small, the heat flux reduces to the classical form,
which includes advection as well as parallel thermal conduction.

To obtain the evolution equations for the electron density and temperature, it is necessary to write Eq.~(\ref{equation:f}) to second order. It takes the form
\begin{equation}
   \frac{\partial}{\partial t}f_1+v_\parallel\nabla_\parallel f_1-\frac{e}{m_e}E_\parallel\frac{\partial }{\partial v_\parallel}f_1=C_0(f_2)+C_1(f_1)+C_2(f_0),
  \label{equation:f2}
\end{equation}
where $C_0$ and $C_1$ are given in Eqs.~(\ref{C0}) and (\ref{C1}). There are
terms from the second order collision operator $C_2$ that arise from the
velocity dependence of the collision rates $\nu_{ei}$ and
$\nu_w$. However, these terms all cancel when computing the continuity
and energy moments of Eq.~(\ref{equation:f2}) so the minimal required
expression for $C_2$ is given by
\begin{equation}
  \begin{aligned}
  C_2=  & \frac{1}{2}\nu_w(v)\frac{\partial}{\partial {\bf v}}\cdot\left( \vec{\vec{I}}(U+V_w)^2+({\bf U}+{\bf V}_w)({\bf U}+{\bf V}_w)\right)\cdot\frac{\partial }{\partial {\bf v}}\\
 &  +(\nu_w\rightarrow\nu_{ei}, {\bf V}_w\rightarrow 0).
  \label{C2}
  \end{aligned}
\end{equation}
The evolution equation for the plasma density is obtained by
integrating Eq.~(\ref{equation:f2}) over the velocity. All of the
collision terms on the right side of the equation as well as the
$E_\parallel$ term integrate to zero, leaving the continuity equation
as written in Eq.~(\ref{continuity}).  Note that it is not necessary
to evaluate $f_2$ to obtain the density evolution equation since
$C_0(f_2)$ integrates to zero. The equation for the evolution of the
electron energy is obtained by multiplying Eq.~(\ref{equation:f2}) by
$m_ev^2/2$ and integrating over the velocity. As in the
continuity equation, $f_2$ drops out since in a stationary frame where
$C_0$ is evaluated the scattering does not change the particle
energy. The electron energy evolution equation takes the form
presented in Eq.~(\ref{energy_electrons}).

As discussed in Sec.~\ref{derivation}, the temperature gradient drive of the whistlers is only nonzero if the whistler scattering rate is energy
dependent.  The energy dependence of the whistler-driven scattering
rate was not evaluated in PIC simulations of heat flux driven
whistlers \citep{Roberg-Clark16,Roberg-Clark18,Komarov18} but the
bounce frequency of electrons trapped in large-amplitude whistlers as
seen in the simulations increases with the particle energy
\citep{Karimabadi90}. However, in real systems in which the temperature
gradient scale length is large, the whistler wave amplitude will be small (see discussion in Sec.~\ref{derivation}) and with a broader
spectrum of waves than documented in the PIC simulations. The
quasilinear scattering rate for an assumed spectrum of low-frequency
Alfv\'en waves has been calculated previously
\citep{Lee82,Schlickeiser89,Schlickeiser98}. The velocity dependence of
the scattering rate takes the form of a powerlaw $v^\gamma$ where $\gamma = q-1$
and where the wave energy spectrum was assumed to have a powerlaw form
$k^{-q}$. Since the whistler waves of interest in electron scattering
are also sub-cyclotron these earlier quasilinear results also apply to the electrons'
response to whistlers. The spectrum of whistler waves in PIC
simulations of heat flux driven whistlers fell off steeply
\citep{Roberg-Clark18} but the limited spectral range of the
simulations likely impacted the results. The cascade of electron-MHD
turbulence has also been explored \citep{Biskamp99} and yielded
powerlaw spectra that scaled like $k^{-7/3}$ so that $\gamma\sim 4/3$. Thus, we take the whistler scattering rate to take the
quasilinear form with a powerlaw dependence on velocity:
\begin{equation}\nu_{w}=0.1\Omega_{e}\frac{\varepsilon_w}{\varepsilon_B}(v/v_{te})^\gamma,
  \label{nu_ql}
\end{equation}
with $\gamma\sim 4/3$. The numerical factor of $0.1$ in
Eq.~(\ref{nu_ql}) is based on the results of PIC simulations \citep{Roberg-Clark18} rather than a detailed
quasilinear calculation. However, as shown in Sec.~\ref{overview},
while this factor impacts the rate of growth of whistlers, it does not
control their saturation and the associated saturated value of
$\nu_{we}$. Since the time scale for whistler growth is short compared
with dynamical timescales, which vary as $L/V_w$, the details of the
whistler growth rate are not important. More important is the onset
criterion for whistler growth, which, as shown in Sec.~\ref{overview},
is independent of the factors appearing in Eq.~(\ref{nu_ql}).
\subsection{Whistler wave equation}
Finally, we present a derivation of  Eq.~(\ref{energy_whistlers}), which
describes the whistler energy transport. The extraction and heating
terms follow from the transfer of energy between the electrons and
whistlers through the scattering process. The calculation that leads
to the advection terms on the left side of the equation, which are
important in describing the propagation of the whistlers and
evaluating their interaction with the bulk fluid, is presented
here. However, to simplify the equations we limit the calculation to a simple system with straight magnetic field lines. The more general case is presented in the Appendix. We start with Faraday's law for the perturbed whistler
magnetic field,
\begin{equation}
  \frac{\partial}{\partial t}{\bf \delta B}_w+c\nabla\times{\bf \delta E}_w=0,
  \label{faraday}
\end{equation}
where $\delta \mathbf{B}_w$ and $\delta \mathbf{E}_w$ are the transverse magnetic and
electric field of the whistler wave. Although the scattering of
electrons by the whistlers requires that the waves be oblique with
respect to the ambient magnetic field, here we explore only the transport
along the ambient magnetic field and therefore consider a simple
1D model of the wave dynamics. Since the full 3D MHD equations will be displayed in the Appendix, the convection velocity ${\bf U}$ of the MHD fluid is taken as a general vector. Taking the dot product of Eq.~(\ref{faraday}) with ${\bf \delta B}_w$ and completing some vector algebra, we obtain an equation for the wave energy $\varepsilon_w=|{\bf \delta B}_w|^2/8\pi$
\begin{equation}
  \frac{\partial}{\partial t}\varepsilon_w+\frac{c}{4\pi}{\bf \nabla}\cdot ({\bf \delta E}_w\times {\bf\delta B}_w)+{\bf\delta J}_w\cdot {\bf\delta E}_w= -V_w\left(\alpha_{10}\frac{\nu_{we}}{\nu_e}n\nabla_\parallel T+F_w\right),
  \label{energy_whistler1}
\end{equation}
where we have used Amp\`ere's law without the displacement current and have included the scattering terms on the right-hand side. 
To evaluate ${\bf\delta J}_w\cdot {\bf\delta E}_w$, we first note that ${\bf\delta J}_w=-ne{\bf\delta v}_w$ and 

An equation for the whistler electric field ${\bf E}_w$ is obtained from the linearized electron equation of motion with no inertia
\begin{equation}
  0={\bf\delta E}_w+\frac{1}{c}{\bf\delta v}_w\times {\bf B}+\frac{1}{c}{\bf U}\times {\bf\delta B}_w,
  \label{momentum}
  \end{equation}
where the electron streaming velocity ${\bf U}$ is equal to that of the ions because of the current in an MHD systems is zero unless Hall terms in Ohm's law are retained \citep{Drake19,Arnold19}. In the Poynting flux the whistler electric field ${\bf\delta E}_w$ takes the form,
\begin{equation}
  {\bf\delta E}_w=-\frac{1}{c}({\bf U}+{\bf V}_w)\times {\bf\delta B}_w.
  \label{electricfield}
\end{equation}
This is equivalent to transforming from the whistler wave frame, where the electric field is zero, into the laboratory frame. The Poynting flux takes the form
\begin{equation}
    {\bf S}_w=\frac{c}{4\pi}{\bf\delta E}\times {\bf\delta B}=\left( 2(U_\parallel +V_w){\bf b}+{\bf U}_\perp\right)\varepsilon_w.
    \label{poynting}
\end{equation}
Using Eq.~(\ref{momentum}) to calculate ${\bf\delta E}_W$, the heating term can be written as ${\bf\delta J}\cdot {\bf\delta E}_w={\bf U}\cdot {\bf F}_{pw}$ where ${\bf F}_{pw}$ is the whistler ponderomotive force acting on the electrons, which is given by
\begin{equation}
    {\bf F}_{pw}=-\frac{ne}{c}{\bf\delta v}_w\times {\bf\delta B}_w.
\end{equation}
The right side is easily evaluated using Amp\`ere's law for ${\bf v}_w$ and carrying out some vector algebra. The resulting expression for the ponderomotive force is
\begin{equation}
{\bf F}_{pw}=-{\bf bb}\cdot {\bf\nabla }\varepsilon_w
\label{Fpw}
\end{equation}
Inserting ${\bf F}_{pw}$ and ${\bf S}_w$ into
Eq.~(\ref{energy_whistler1}) yields the transport equation for
$\varepsilon_w$ given in Eq.~(\ref{energy_whistlers}). The interaction
between the whistler radiation pressure and the ions through the last
term on the left side of the whistler transport equation requires a
corresponding term in the ion momentum equation. This arises when the
individual electron and ion momentum equations are added to produce
the one-fluid momentum equation. For completeness the full set of MHD equations along with the
whistler constrained electron energy transport equations are presented
in the Appendix. The results in the Appendix generalize the ponderomotive force to the case when magnetic fields have curvature. 

\section{Simulations of electron thermal transport with whistler scattering}
\label{computations}
While the basic characteristics of whistler limited transport have
been discussed in Sec.~\ref{overview}, here we show the results of
numerical solutions of the coupled equations for the electron
temperature $T_e$ and the whistler energy density
$\varepsilon_w^\pm$. For simplicity, we freeze the ions so that the
bulk flow $U$ is zero and the plasma density remains constant. In the
first test case we consider a system with an initial temperature
profile and temperatures at the
boundaries specified such that the temperature gradient is negative.  We assume zero slope boundary conditions on
so that the boundary temperatures can float. The
whistler waves are assumed to have small amplitude at $t=0$ so that
classical scattering dominates the early evolution. Because the
temperature gradient is zero at the boundaries, the threshold for
whistler growth in Eq.~(\ref{growthrate_threshold}) is not satisfied close to the boundaries
so the whistler wave amplitude remains small there. As a
consequence, there is little heat flux through the boundaries and the integrated total energy (electron plus whistlers)
in the system is conserved. Because the gradient of the
temperature in the system is zero or negative everywhere, only
rightward propagating waves are included since leftward propagating
waves are damped. 

It is convenient to normalize the equations to reduce the number of
free parameters. The temperature is normalized to its value on the left boundary, $T_0$,
and the whister wave energy to $n_0T_0$. Lengths are normalized to the
length of the computational domain $L$ and time to the transit time of
the whistler across the system $L/V_{w0}$ with $V_{w0}$ based on the
parameters $T_0$ and $\beta_0$ of the left boundary. The scattering
rate $\nu_e$ is normalized to the nominal saturated whistler
scattering rate $\beta_0v_{te0}/L$ written below
Eq.~(\ref{nu_whistler}). The resulting normalized coupled equations for $T$ and $\varepsilon_w^\pm$ take the simple form
\begin{equation}
\frac{3}{2}\frac{\partial}{\partial t}T+\nabla_\parallel Q=(\tilde{V}_{\mathrm{st}}^+-\tilde{V}_{\mathrm{st}}^-)\nabla_\parallel T +H_w^++H_w^-
  \label{T_n}, 
  \end{equation}
where for notational simplicity we have not relabeled the variables ({\it e.g.}, $T/T_0\Rightarrow T$). In Eq.~(\ref{T_n})
\begin{equation}
    Q=\left( \tilde{V}_{\mathrm{st}}^+-\tilde{V}_{\mathrm{st}}^-\right) T-\kappa_e\nabla_\parallel T,
    \label{Q_n}
\end{equation}
with $\kappa_e=T/\nu_e$, $\nu_e=\nu_{ei}+\nu_{we}^++\nu_{we}^-$ and 
\begin{equation}
    \nu_{ei}=\left(\frac{L}{\beta_0L_{c0}}\right)\frac{1}{T^{3/2}}, \;\; \nu_{we}^\pm=0.1\varepsilon_w^\pm\frac{L}{\rho_{e0}}. 
      \label{nu_n}
  \end{equation}
The streaming velocities $\tilde{V}_{\mathrm{st}}^\pm$ are defined as
\begin{equation}
  \tilde{V}_{\mathrm{st}}^\pm =V_w\frac{\nu_{we}^\pm}{\nu_e},
    \label{Vst}
\end{equation}
and $V_w=T^{-1/2}$. The heating functions take the form
\begin{equation}
  H_w^\pm =V_w^2\left( \nu_{we}^\pm+\frac{\nu_{we}^+\nu_{we}^-}{\nu_e}\right).
    \label{Hw}
\end{equation}
The equation for the whistler wave energy ($\varepsilon_{we}^\pm/n_0T_0\Rightarrow\varepsilon_w^\pm$) is
\begin{equation}
  \frac{\partial }{\partial t}\varepsilon_w^\pm \pm 2\nabla_\parallel V_w\varepsilon_w^\pm=
  \mp \tilde{V}_{\mathrm{st}}^\pm\nabla_\parallel T-H_w^\pm.
  \label{E_wn}
\end{equation}
 The equations have two important free parameters, the collisionality
 parameter $L/\beta_0L_{c0}$, which for values less than unity, leads
 to whistler growth and the Larmor radius parameter $L/\rho_{e0}$
 which controls the rate of growth of whistlers compared with the
 whistler transit time across the domain -- a large value of $0.1
 L/\rho_{e0}$ indicates that in the unstable domain the whistlers will
 reach large amplitude in a time short compared with the global
 evolution of the system. For simplicity, we have set the various
 parameters $\alpha_{i}$ to unity since their values do not significantly impact the dynamics.

 For computational reasons we do not actually evolve the whistler wave
 energy as written in Eq.~(\ref{E_wn}). The equation leads to numerical
 problems in spatial locations where $\varepsilon_{w}^\pm\Rightarrow 0$ because
 the time step can cause $\varepsilon_w$ to become negative. However,
 both of the terms on the right side of the equation are proportional
 to $\nu_{we}^\pm$ and therefore to $\varepsilon_w^\pm$. Thus, the equation
 can be divided by $\varepsilon_w^\pm$ and the equation can be written as
 an evolution equation for $\ln\varepsilon_w^\pm$. This is in effect a
 stretching transformation for $\varepsilon_w^\pm$ around zero. 
 The coupled Eqs.~(\ref{T_n}) and (\ref{E_wn}) are solved with the 1D Galerkin/Petrov-Galerkin method \citep{Skeel90} using MATLAB.
 
\begin{figure}
\centering
\includegraphics[keepaspectratio,width=6.5in]{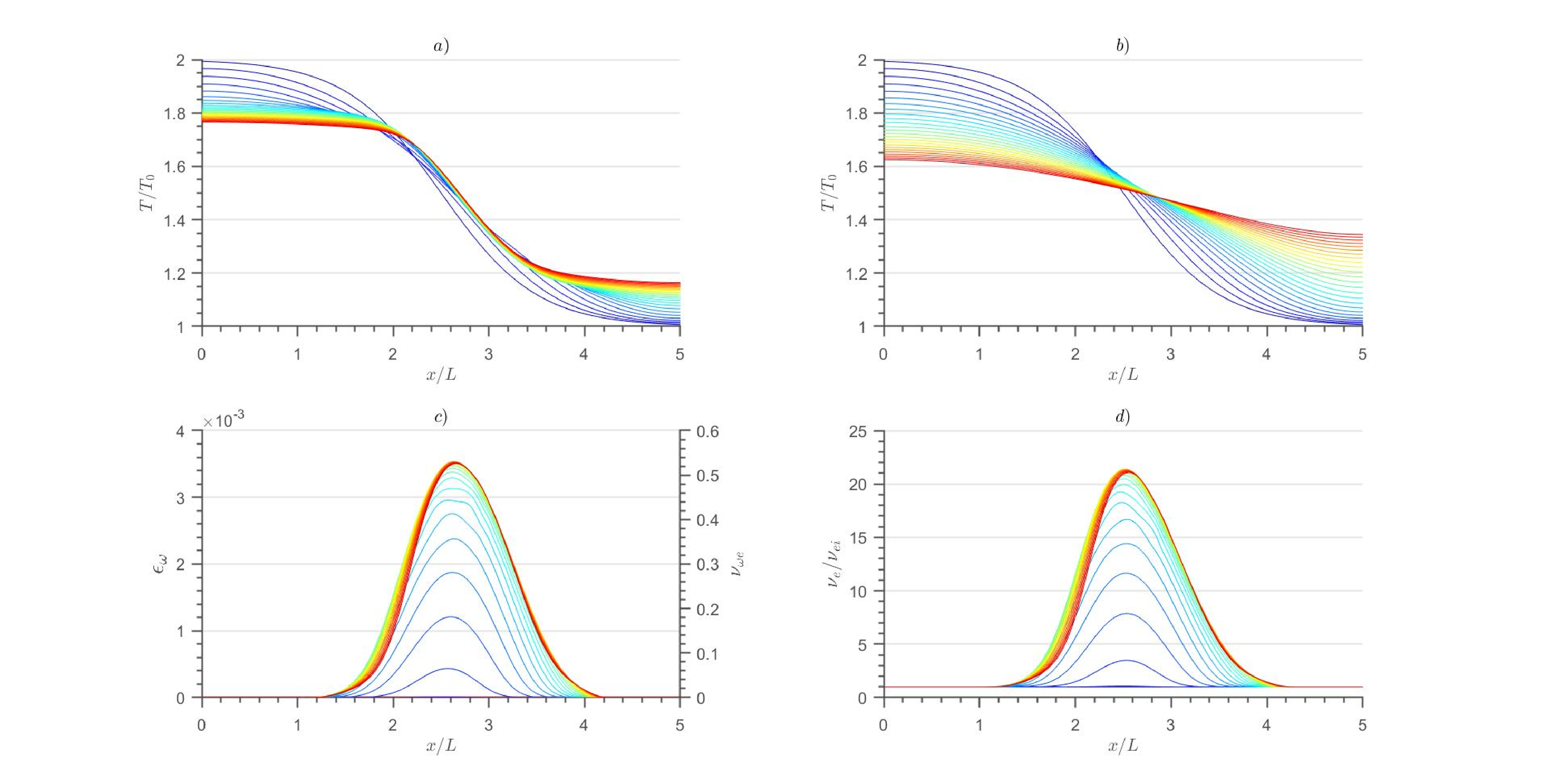}
\caption{Results of a simulation of electron thermal transport in a
  regime with unstable whistlers ($L/\beta_eL_c=0.05$). At several times
  ($t\in (0,0.15)$) during the growth phase of whistlers, cuts of the
  electron temperature $T$ with whistler scattering included (a) and
  with whistler scattering eliminated (b). From the simulation in (a)
  cuts of the energy density of rightward propagating whistlers
  $\varepsilon_w^+$ and the associated whistler scattering rate
  $\nu_{we}^+$ in (c) and the ratio of the total electron scattering
  rate $\nu_e=\nu_{ei}+\nu_{we}^+$ to the electron-ion scattering rate
  $\nu_{ei}$ in (d). }
\label{tanh_early}
\end{figure}

 In Fig.~\ref{tanh_early} we show cuts of (a) the temperature profile,
 (c) the whistler wave energy profile $\varepsilon_w^+$ and associated
 whistler scattering rate $\nu_{we}^+$ and (d) the profile of the
 ratio of $\nu_e=\nu_{ei}^e+\nu_{we}^+$ to $\nu_{ei}^e$ at several
 times during the growth phase of the whistler waves ($t\in
 (0.0.15)$). The leftward propagating whistler is stable for this
 simulation since $\nabla_\parallel T$ is never positive. The parameters for this simulation were
 $L/\beta_0L_{c0}=0.05$, $L/\rho_{e}=1500$ with the initial temperature
 profile given by $1.5-0.5\tanh (x-2.5)$. At the
 beginning of the evolution when $\nu_{we}^+$ is small,
 $\nu_e=\nu_{ei}\sim L/\beta_0L_{c0}=0.05$ so $\kappa\sim 20$. The initial 
 temperature evolution is rapid. However, the whistlers also grow
 rapidly until $\nu_e\sim 1$ and the evolution of the temperature
 slows dramatically. For comparison, the evolution of the temperature
 is shown in (b) with the whistler wave energy set to zero,
 corresponding to classical transport. The flattening of the
 temperature takes place much more rapidly in the absence of the
 suppression of transport by whistlers.

\begin{figure}
\centering
\includegraphics[keepaspectratio,width=6.5in]{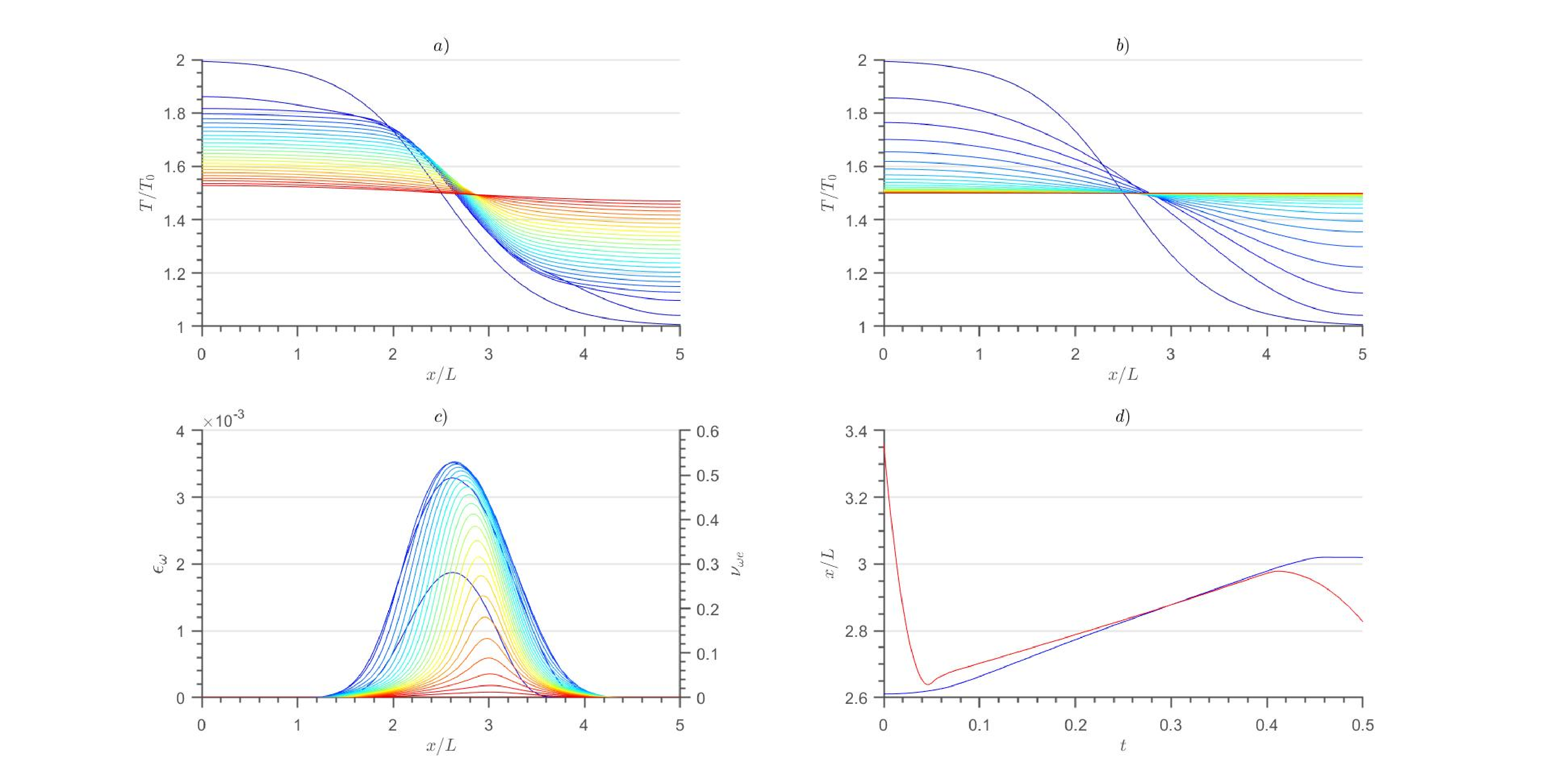}
\caption{Results of the simulation of Fig.~\ref{tanh_early} over a longer time interval ($t\in (0,1.5))$. Cuts in (a), (b) and (c) as in Fig.~\ref{tanh_early}. In (d) the trajectory of the location of the peak of $\varepsilon_w^+$ (blue) and the maximum temperature gradient (orange). }
\label{tanh_late}
\end{figure}

 The time evolution of the temperature during the phase when whistler
 growth has saturated ($t\in (0,1.5)$) is shown in
 Fig.~\ref{tanh_late}(a). The plots in (a), (b) and (c) are as in
 Fig.~\ref{tanh_early}. During this time the profiles of the whistler
 wave amplitude and scattering rate $\nu_{we}$ evolve slowly,
 remaining in the saturated state defined by
 Eq.~(\ref{nu_whistler}). That the temperature is advected by the
 whistlers can be seen by the trajectories of the location of the peak of the temperature gradient and of the whistler
 energy spectrum shown in (d). Both peaks propagate together because the whistlers rapidly reach equilibrium with the local temperature gradient as given in Eq.~(\ref{nu_whistler}). The peaks propagates with nearly
 constant velocity of order unity, which corresponds to $V_w$ in our
 normalized system.

\begin{figure}
\centering
\includegraphics[keepaspectratio,width=6.5in]{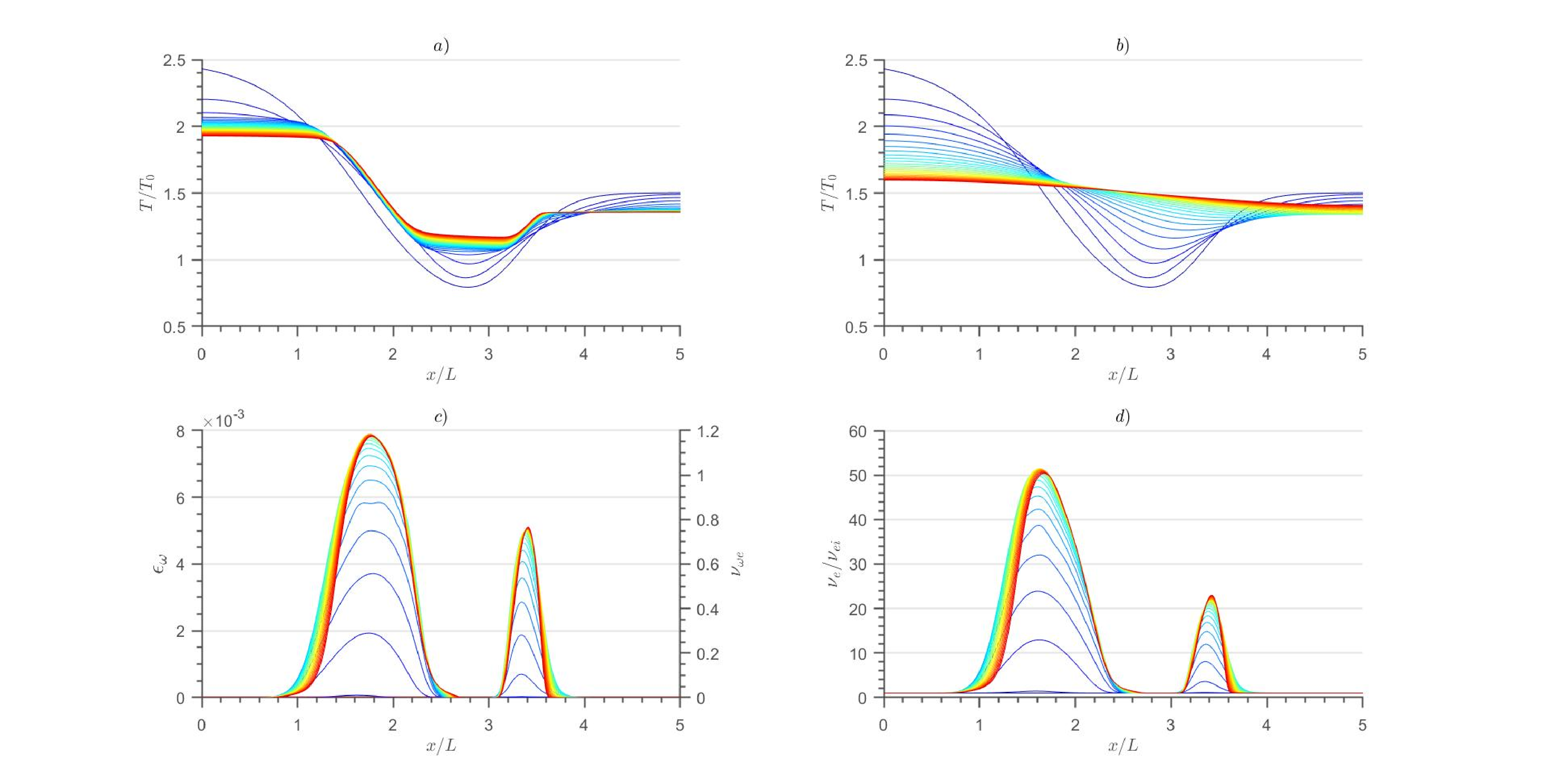}
\caption{Results of a simulation of electron thermal transport in a system 
 with a bi-directional temperature gradient and unstable whistlers propagating in the positive and negative directions. At several times ($t\in
  (0,0.1)$) during the growth phase of whistlers, cuts of the electron
  temperature $T$ with whistler scattering included in (a) and with
  whistler scattering eliminated in (b). From the simulation in (a) cuts
  of the energy density of rightward,  $\varepsilon_w^+$, and leftward  $\varepsilon_w^-$, propagating whistlers and the associated whistler scattering rates
  $\nu_{we}^\pm$ in (c).  In (d) the ratio of the total electron scattering
  rate $\nu_e=\nu_{ei}+\nu_{we}^++\nu_{we}^-$ to the electron-ion scattering rate
  $\nu_{ei}$. }
\label{bidirection_early}
\end{figure}

 We now consider a more complex initial temperature profile with
 regions of positive and negative temperature gradient so that whistlers
 propagating to the left and right $\varepsilon_w^\pm$ must be
 included. The parameters for this simulation were the same as in
 Fig.~\ref{tanh_early} but with an initial temperature profile given
 by $2.0-\tanh (x-5/3)+0.5\tanh\left( (x-10/3)/0.5\right)$. As in the
 simulation of Fig.~\ref{tanh_early}, we show in 
 Fig.~\ref{bidirection_early} cuts of (a) the temperature
 profile, (c) the whistler wave energy profiles $\varepsilon_w^\pm$ and
 associated whistler scattering rates $\nu_{we}^\pm$ and (d) the
 profile of the ratio of $\nu_e=\nu_{ei}^e+\nu_{we}^++\nu_{we}^-$ to
 $\nu_{ei}^e$ at several times during the growth phase of the whistler
 waves ($t\in (0.0,0.1)$). As in the previous simulation, the temperature
 evolution is initially rapid. However, the whistlers grow rapidly
 with rightward propagating whistlers growing in the region of
 negative temperature gradient and the leftward propagating whistlers
 on the positive temperature gradient. Once $\nu_e\sim 1$ the
 evolution of the temperature slows dramatically. For comparison, the
 evolution of the temperature is shown in (b) with the whistler wave
 energy set to zero, corresponding to classical transport. Again the
 flattening of the temperature takes place much more rapidly in the
 absence of the suppression of transport by whistlers. An important
 result from the simulation with bidirectional whistlers is the
 absence of a region of significant overlap of the oppositely propagating
 waves. This is a consequence of the threshold for whistler growth
 given in Eq.~(\ref{growthrate_threshold}). For a shallow temperature
 gradient classical collisions prevent whistler onset so there is
 always a stable band between the regions where the right and left
 going waves are unstable. The consequence is that two classes of
 waves develop in spatially distinct regions of space. This differs
 from the case of cosmic ray transport limited by Alfv\'en waves where
 counter-streaming waves can develop provided the wave damping rate is sufficiently low.
 
\begin{figure}
\centering
\includegraphics[keepaspectratio,width=6.5in]{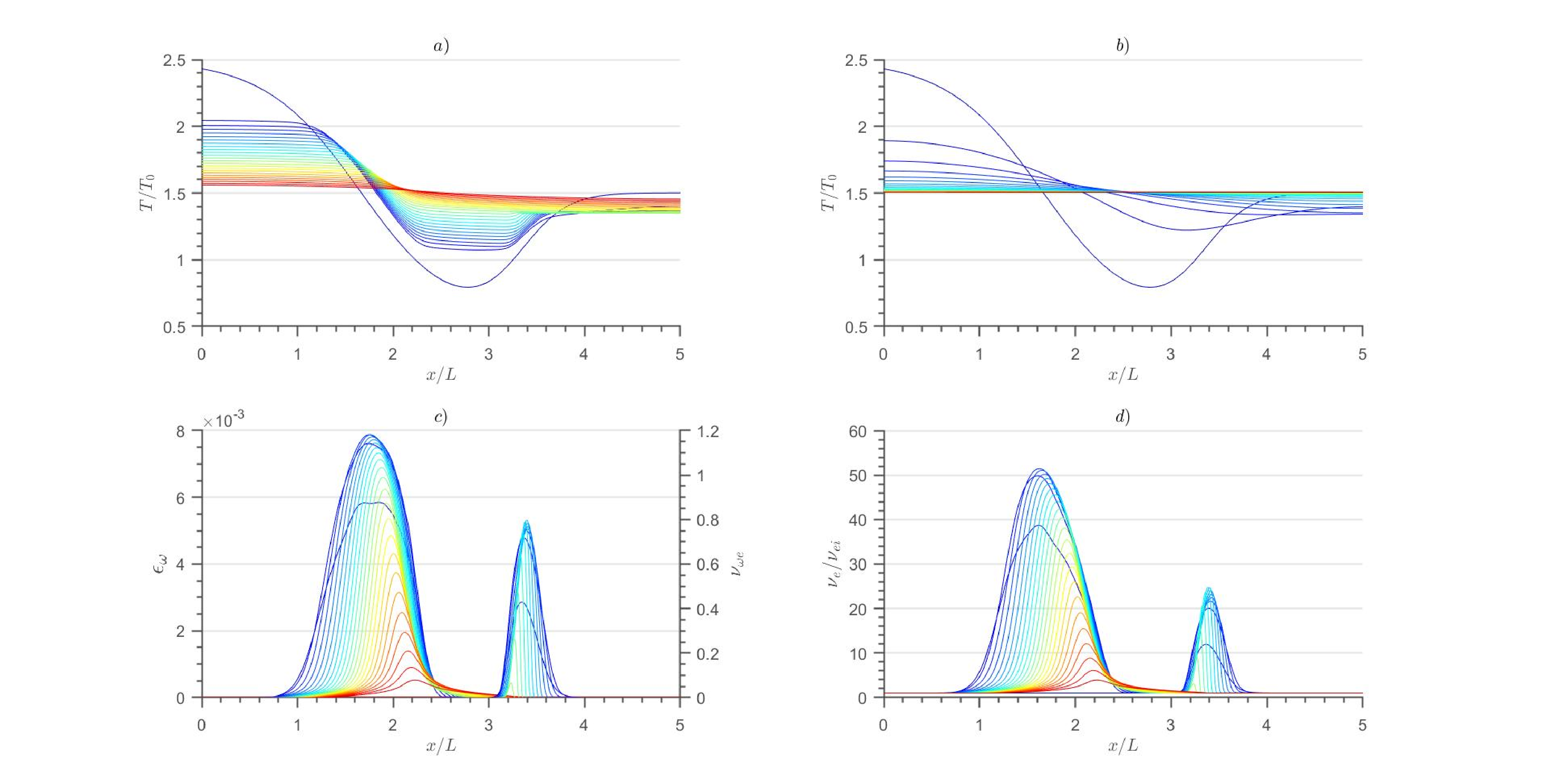}
\caption{Results of the simulation of Fig.~\ref{bidirection_early} over a longer time interval ($t\in (0,0.5))$. Cuts in (a), (b), (c) and (d) as in Fig.~\ref{bidirection_early}. }
\label{bidirection_late}
\end{figure}

\begin{figure}
\centering
\includegraphics[keepaspectratio,width=6.5in]{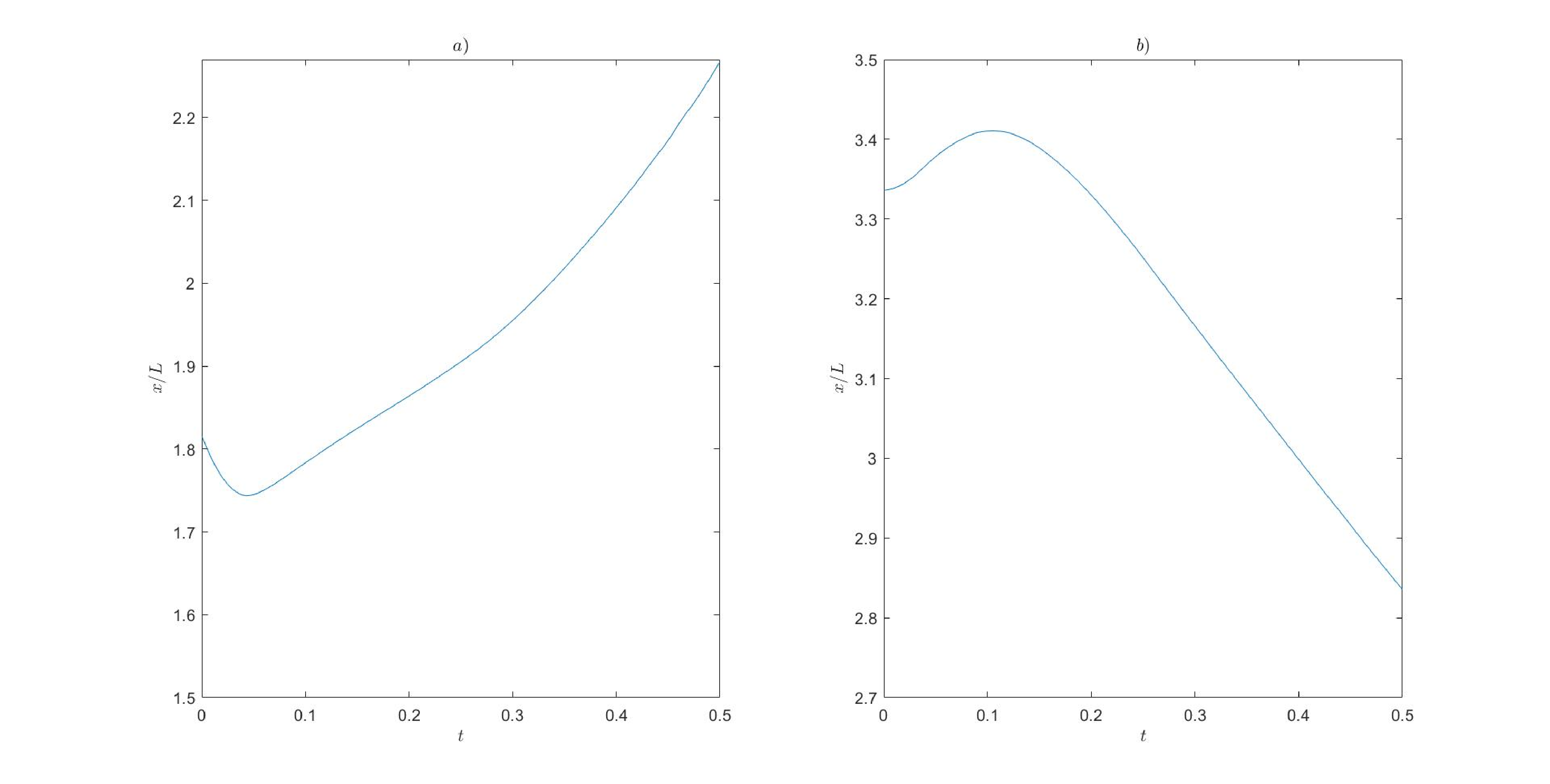}
\caption{Results from the simulation of Fig.~\ref{bidirection_late}, the trajectories of the peaks of the wave energy of the rightward (a) and leftward (b) propagating whistlers. }
\label{bidirection_traj}
\end{figure}

 In Fig.~\ref{bidirection_late} we show the late time evolution of the
 profile shown in Fig.~\ref{bidirection_early}. In (a) is the whistler
 limited profile evolution and in (b) the classical transport
 result. In the case of whistler limited transport the temperature
 minimum fills in as the temperature gradients on the left and right
 propagate towards the middle to fill in the temperature dip. This
 behavior is perhaps even more evident in (c), which displays the
 evolution of the whistler energy spectra. The peaks of the spectra
 move towards the temperature minimum, following the location of the
 maximum temperature gradient.  $\varepsilon_w^+$ moves to the right
 and $\varepsilon_w^-$ moves to the left. In
 Fig.~\ref{bidirection_traj}, the trajectory of the peak of the
 spectrum of $\varepsilon_w^+$ is shown in (a) and that of
 $\varepsilon_w^-$ is shown in (b). Again, the propagation velocities
 are of order unity and reflect the motion of the location of the
 maximum temperature gradient.

\section{Discussion and Conclusions}
\label{conclusions}
A set of transport equations for electron energy that includes both
classical electron-ion collisions and self-consistent scattering by
whistler waves has been developed. The whistlers are driven unstable
by the electron temperature gradient along the ambient magnetic
field. For temperature scale lengths below the critical value
$L=\beta_eL_c$, with $L_c$ the classical electron mean-free-path and
$\beta_e$ the ratio of electron to magnetic pressure, whistler waves
will rapidly grow and reduce electron transport below that based on
the classical Spitzer conductivity. For typical values of $L_c$ in the
ICM ranging from $0.1-1$~kpc and $\beta_e\sim 100$, large regions of the ICM
are likely to be controlled by whistler-limited thermal transport.

While the impact of whistler constrained thermal transport on the dynamics of the ICM will require detailed calculations beyond the scope of the present paper, simple scaling arguments suggest the broad importance of the results for the ICM. In the following, we will show that (1) the suppression of thermal conduction decreases the characteristic length below which thermal instability is suppressed \citep{Field65}, (2) the temperature dependence of heat flux transitions from $T_e^{7/2}/L$ to $V_wnT_e\sim T_e^{1/2}$ as a system transitions to the regime where classical collisions are weak, (3) the magneto-thermal (MTI) and heat-flux-driven buoyancy (HBI) instabilities are
impaired or completely inhibited by whistler limited heat flux, and (4) the heat flux constraint allows sound waves, which are significant players in heating the cool cores of the ICM plasma, to propagate greater distances from the central black hole, consistent with observations. Finally, we discuss the potential of direct measurements in the high $\beta$ solar wind to validate the new model.

\subsection{The impact on thermal condensation}\label{thermalinstability}

Further constraints on electron thermal transport from whistler scattering will make it even
less likely that heat conduction from the outer regions of the ICM to
the cool core can limit the radiative collapse at the largest
scales. On the other hand, an important question is whether the local
dynamics of radiative instabilities that are likely to develop within
the cool core will be altered compared with the classical
model. In the case of an isobaric perturbation about an equilibrium in which radiative cooling is balanced by some local heat source, the suppression of thermal transport relative to the classical Spitzer value will decrease the Field length \citep{Field65}, thereby extending the domain of thermal instability down to smaller spatial scales.  Further, the significant  consequences of whistler limited transport can be more clearly identified by considering a state in which radiative cooling is balanced by a non-zero divergence of the heat flux.  In the simplest one-dimensional model of an isobaric condensation, the heat flux in the classical model
scales as $T^{7/2}/L$ with $L$ the ambient temperature scale
length. To maintain a constant heat flux into the condensing region
where the temperature is dropping requires the local temperature
gradient to increase or $L$ to decrease. The reduction in $L$ pushes the system towards the whistler unstable domain. Specifically, the heat flux $Q$ scales as
\begin{equation}
  Q\sim\frac{L_c}{L}nT_ev_{te}
\end{equation}
so the whistler instability criterion can be written as
\begin{equation}
  \frac{L}{\beta_eL_c}\sim\frac{nT_ev_{te}}{\beta_eQ}\sim\frac{B^2}{8\pi Q}v_{te}<1.
  \label{whistleronset}
  \end{equation}
Thus, in a 1D system where $B$ is a constant, the threshold for
whistler onset will be crossed when the temperature is low enough. At
that point the scaling of the heat flux will change from $T^{7/2}/L$
to $V_wnT_e\sim T_e^{1/2}$. Thus, the dynamics of condensation is
likely be substantially changed in light of the new transport model.

The temperature gradients across cold fronts can be quite large and,
depending on the strength of the locally draped magnetic field, are
likely to drive whistlers that in turn can control transport and the
structure of these fronts. The equations developed here provide a
fully self-consistent framework for exploring the dynamics and
structure of cold fronts with no constraints on the possible breakdown
of the classical model of thermal transport.

\subsection{The impact of heat flux suppression on anisotropic transport instabilities}

A pure hydrodynamic and gravitationally stratified plasma is unstable to perturbations that cause convective motions if the entropy is a decreasing function of radius, i.e., if $\partial \ln K/\partial \ln r < 0$  \citep{Schwarzschild1906}, where $K=P\rho^{-\Gamma}$ is the entropic function labeling an adiabatic curve, $\rho$ is the mass density, and $\Gamma=5/3$ is the ratio of specific heats. The entropy profiles of all observed galaxy clusters are increasing functions of radius, which should render them convectively stable \citep{Cavagnolo08}. If a fluid element in a stably stratified atmosphere is adiabatically displaced in radius from its equilibrium position by $\delta r$, it experiences a buoyant restoring force per unit volume \citep{Ruszkowski10}
\begin{align}
\label{eq:F_adiab}
F_\mathrm{adiab}=\rho\ddot{r}= -\frac{\rho g}{\Gamma} \frac{\partial \ln K}{\partial r} \delta r,
\end{align}
where $g$ is the gravitational acceleration, that causes oscillations around its equilibrium position at the classical Brunt-V\"ais\"al\"a frequency $N$ where
\begin{eqnarray}
   N^2 = \frac{g}{\Gamma r}\frac{\partial \ln K}{\partial \ln r}.
\end{eqnarray}
More fundamentally, the Brunt-V\"ais\"al\"a frequency is the limiting frequency of linear internal gravity waves (which have frequency $\omega_{\rm gw}^2=k^2_h N^2/|k^2|$ where $k_h$ is the component of the wavenumber orthogonal to the gravity gradient), so real $N$ translates into linear stability of the atmosphere. 
However, if an external turbulent driving force is larger than the buoyant restoring force then it can overcome the stable equilibrium and induce mixing.

This paradigm experiences a fundamental change in the weakly collisional, magnetized plasma of a galaxy cluster because it changes the response of the plasma to perturbations. Provided there is a radial temperature gradient, $\partial T_e / \partial r \ne 0$, anisotropic thermal conduction along the mean magnetic field causes the ICM to be almost always buoyantly unstable regardless of the sign of the temperature and entropy gradients. If $\partial T_e / \partial r > 0$ (which applies to the central cooling regions in cool core clusters) the ICM is unstable to the heat-flux-driven buoyancy instability \citep[HBI,][]{Quataert08} provided that there are regions where the magnetic field is mostly radially aligned. If $\partial T_e / \partial r < 0$ (which applies to all clusters on large scales and is a consequence of hydrostatic rearrangement in the presence of the universal dark matter potential) the ICM is unstable to the magneto-thermal instability \citep[MTI,][]{Balbus00} provided that there are regions where the magnetic field is mostly horizontally aligned. Using a linear stability analysis \citep{Kunz2011} and non-linear simulations \citep{Kunz2012} it was shown that anisotropic viscosity (i.e., Braginskii pressure anisotropy) effects the MTI and HBI growth rates by stifling the convergence/divergence of magnetic field lines, but cannot suppress it. When field geometries are choosen such as to stabilize the HBI/MTI, the system is still subject to related overstabilities \citep{Balbus10}. In consequence, any perturbation will be convectively unstable and cause instant mixing of the thermal plasma because it lacks a restoring force that provides an energy penalty \citep{Sharma09} which facilitates advection and turbulent mixing of AGN injected energy \citep{Kannan17}. In the saturated state of these anisotropic transport instabilities, the buoyant restoring force is altered to $|F_\mathrm{cond}|\sim \rho g (\partial\ln T_e/\partial r)\delta r$, and replaces the restoring force of Eq.~\eqref{eq:F_adiab} that is based on the classic Schwarzschild criterion \citep{Sharma09}.

Kinetic plasma physics and finite Larmor-radius effects can however significantly impact those HBI/MTI instabilities, which have been found using a magnetofluid (i.e.\ Braginskii) description. While at long wavelengths (the `drift-kinetic' limit), a kinetic analysis reinforces the MTI and its Alfv{\'e}nic counterpart, at sub-ion-Larmor scales, there is an overstability driven by the electron-temperature gradient of kinetic-Alfv{\'e}n drift waves whose growth rate is even larger than the standard MTI \citep{Xu2016}. While the effective heat conductivity can locally be supressed by the ion-scale mirror instability \citep{Komarov16,Riquelme16}, recent work found that the nonlinear saturation of the MTI is not significantly modified \citep{Berlok2021}. Given that the predicted reduction in parallel thermal transport can be significant in our picture of whistler-mediated thermal conduction, an important question is whether this reduction might affect these large-scale MTI and HBI fluid instabilities. Formally, the current treatments of MTI and HBI explicitly assume diffusive parallel heat transport and so would need to be reworked with our modified scheme that highlights the importance of advective transport.  However, as a first approach, it is possible to argue physically.  The MTI/HBI rely on three properties: (i) the presence of a radial temperature gradient, $\partial T_e / \partial r \ne 0$, (ii) the large ratio of electron thermal transport along to that across the ambient magnetic field and (iii) a short conduction time in comparison to the buoyancy and advection time scales on the length scales considered.\footnote{As noted by our referee, there is an additional property: the viscous time cannot be too short in comparison to the buoyancy time. Because the conduction time and the viscous time are related (in Braginskii's MHD), one cannot take the limit of fast conduction while neglecting the anisotropic viscous stress \citep{Kunz2011}, which also suppresses the HBI by stifling the convergence/divergence of magnetic-field lines that is responsible for generating the buoyant motions.}

The perpendicular thermal transport arising
from the whistler instability has not been explored in detail. The
invariance of the canonical momentum in the symmetry direction of a 2D
PIC model constrains particle motion perpendicular to ${\bf B}$ so
that the exploration of perpendicular transport requires the PIC
modeling to be carried out in a more computationally challenging 3D
system. On the other hand, based on the basic characteristics of the
heat-flux-driven whistler, we can estimate the perpendicular
transport. The characteristic transverse scale of the whistler
turbulence is the electron Larmor radius $\rho_e$ so that a reasonable
estimate of the cross-field diffusion is
\begin{equation}
  D_\perp\sim\rho_e^2\nu_{we}.
  \label{Dperp}
\end{equation}
The parallel transport is given by
\begin{equation}
  D_\parallel\sim \frac{v_{te}^2}{\nu_{we}}.
  \label{Dparallel}
\end{equation}
Thus, the ratio $R_\mathrm{trans}$ of the parallel to perpendicular transport is given by
\begin{equation}
  R_\mathrm{trans}=\frac{D_\parallel}{D_\perp}\sim \left(\frac{L}{\beta \rho_e}\right)^2\sim 3\times 10^{26}
  \left(\frac{L}{30\,\mathrm{kpc}}\right)^2
  \left(\frac{T_e}{10^7\,\mathrm{K}}\right)^{-1}
  \left(\frac{B}{1\,\mu\mathrm{G}}\right)^{2},
  \label{Rtrans}
\end{equation}
where we have used the saturated whistler scattering rate
$\nu_{we}=\beta_e v_{te}/L$ given in Eq.~(\ref{nu_whistler}). Thus, the
anisotropy ratio remains extreme, in spite of the factor $\beta_e$
in the denominator of Eq.~(\ref{Rtrans}).
  
While whistler-wave scattering does not qualitatively change the anisotropic character of heat transport and hence the physical basis for the MTI and HBI, the scale $\ell$ that can go unstable is modified. Assuming a hydrostatic atmosphere, these instabilities only grow at interesting rates provided the characteristic time scale at which conduction acts on a given perturbation is much shorter than the buoyancy time scale, which we identify with the inverse Brunt-V\"ais\"al\"a frequency, $N$, and obtain the condition
\begin{align}
  \tau_\mathrm{cond}=\frac{\ell}{V_w}=\frac{\ell \beta_e}{v_{te}}
  \ll N^{-1} = \left(\frac{g}{\Gamma r}\frac{\partial \ln K}{\partial \ln r}\right)^{-1/2}.
\end{align}
This time scale ordering is the basis for the excitation of the MTI and HBI and ensures quasi-isothermality along a given field line \citep{Quataert08}. Because tangled magnetic field lines have a significant portion of azimuthal components not aligned with the vertical buoyancy direction, this adds furthermore to the large separation of time scales. We assume an NFW dark matter density profile \citep{NFW97} which is characterized by a scale radius, $r_s$, and a mass density at the scale radius, $\rho_s$, so that the gravitational acceleration in the inner regions at radii $r<r_s$ is given by 
\begin{align}
  g_0=\frac{G M(<r)}{r^2}=2\pi G\rho_s r_s=\mathrm{const.},
\end{align}
where $G$ is Newton's constant. We furthermore assume an exponentially stratified atmosphere where the pressure scale height is given by $h=v_{tp}^2/g_0$ (where $v_{tp}$ is the isothermal sound speed) so that the critical length scale for instability is
\begin{align}
\label{eq:critial_length}
  \ell \ll \frac{v_{te}}{\beta_e} N^{-1} = \sqrt{\frac{\Gamma}{\Gamma-1}}\frac{v_{te}\,v_{tp}}{\beta_e g_0}\sim 20 
  \left(\frac{\sqrt{T_e T_p}}{10^7\,\mathrm{K}}\right)
  \left(\frac{\beta_e}{100}\right)^{-1}
  \left(\frac{g_0}{10^{-8}\,\mathrm{cm~s}^{-2}}\right)^{-1} \mathrm{kpc}.
\end{align}
On larger length scales, conduction is slower than buoyancy and the system transitions to obey the classical Schwarzschild criterion for convective instability, $\partial\ln K/\partial \ln r<0$. However, on scales smaller than $\ell$, there is no restoring force associated with the entropy gradient and passive scalars such as metals should get easily mixed with the surrounding ICM as explained above. This picture is predicated upon the formal applicability of the MTI and HBI but, for realistic magnetic field strengths in clusters with $\beta_e\sim100$, magnetic tension provides an additional restoring force and can suppress a significant fraction of HBI-unstable modes, thus either impairing or completely inhibiting the HBI on scales smaller than $\sim50-70$~kpc, depending on the unknown magnetic field coherence length and the fraction of volume partitioned with these intermittent strong magnetic fields \citep{Yang16}. These considerations strongly constrain the applicability of the MTI and HBI and call for a careful assessment as to whether they are excited in linear theory at all in the presence of whistler-mediated thermal conduction.

If there are bulk flows in the ICM, the condition of Eq.~\eqref{eq:critial_length} is modified and we require that the conductive time scale is much shorter than the advective time scale for the MTI and HBI to be excited, which can be rewritten into a condition for the advection velocity, 
\begin{align}
\label{eq:critical_U}
  U \ll V_w = \frac{v_{te}}{\beta_e}\sim 100 
  \left(\frac{T_e}{10^7\,\mathrm{K}}\right)^{1/2}
  \left(\frac{\beta_e}{100}\right)^{-1} \mathrm{km~s}^{-1}.
\end{align}
Both conditions of Eqs.~\eqref{eq:critial_length} and \eqref{eq:critical_U} are constraining and may severely limit the applicability of the MTI and HBI in galaxy clusters.
  
\subsection{Acoustic wave dissipation and thermalization of AGN energy in cool cores}
Intermittent activity of bipolar AGN jets at the centers of cool core clusters is expected to generate sound waves in the ICM. Concentric ripples in the X-ray emissivity have been  detected in the Perseus \citep{Fabian2003} and Virgo clusters \citep{Forman2005} and have been interpreted as sound waves generated by central supermassive black holes. These arcs have characteristic widths (or wavelengths) $\lambda$ of order $\sim$1 to 10 kpc and are seen up to several tens of kpc away from cluster centers, suggesting that the waves propagate over substantial fractions of cool core radii before completely dissipating. The recent investigation of \cite{Bambic19} showed that up to 25--30\% of the energy injected as fast jets by the AGN can end up as sound waves, highlighting the relevance of this physics to the question of AGN feedback. \\
\indent
The above argument suggests that sound waves could be a promising agent responsible for converting the mechanical energy of the AGN to the thermal energy of the ICM. Appealing features of this mode of heating are that the waves provide a natural mechanism to: (i) quickly deliver AGN energy to the ICM (i.e., on sound crossing timescales that are typically shorter than radiative cooling timescales), and (ii) distribute the energy over a large fraction of the cool core volume (rather than dissipating the energy close to the AGN jet axis). These are desirable features of the model because they can help to explain why cool cores remain globally thermally stable over times comparable to the Hubble time. 
Furthermore, ICM heating via sound wave thermalization is a gentle process involving very subsonic velocity fluctuations. This is a particularly appealing feature of this mode of heating given recent {\it Hitomi} observations \citep{Hitomi2016,Fabian2017} that suggest that the ICM in the Perseus cluster is very calm. \\
\indent
Tapping of sound wave energy can occur via a number of mechanisms. Hydrodynamical simulations of AGN outbursts invoking Spitzer ion viscosity demonstrated that sound wave dissipation can offset radiative cooling losses and that the waves can propagate significant distances away from cluster centers \citep{Ruszkowski2004a,Ruszkowski2004b}. However, the central cool core regions tend to be somewhat overheated in this model. This problem is further exacerbated by the fact that classical Spitzer conduction is $(m_{p}/m_{e})^{1/2}$ more effective than Spitzer viscosity in dissipating waves. This suggests that transport processes need to be substantially suppressed in the ICM in order to eliminate tension with the observations. The necessary level of suppression was quantified using linear theory by \citet{Fabian2005}, and this work was further extended by \citet{Zweibel2018}, who included the self-limiting nature of dissipation by electron thermal conduction, electron-ion non-equilibration effects, and provided estimates of kinetic effects by comparing to a semi-collisionless theory. \\
\indent
One of the main limitations of the above investigations was that the level of transport was quantified by specifying {\it ad hoc} Spitzer suppression factors. Our model allows one to relax these assumptions and it can be applied to make physically-motivated predictions for the evolution of the dissipating sound waves. In particular, our model self-consistently bridges the transition from the collisional to whistler-mediated transport regimes. Properly accounting for this transition may prove crucial for our understanding of the thermalization of the AGN-induced sound waves — while the sound wave dissipation at the very centers of cool cores may occur via collisional processes, collisionless processes are likely to dominate over a wide range of distances away from cool core centers where $L_{c}\beta > \lambda$. 
For example, the Coulomb mean free path is $L_{c}\sim 2\times 10^{-3}$ kpc and $\sim 4\times 10^{-2}$ kpc at the centers of the Virgo and Perseus clusters, respectively; and at the distance of $\sim 50$ kpc from the center, the corresponding values are $L_{c}\sim 0.4$ kpc and $\sim 0.2$ kpc (see, e.g., \citet{Zhuravleva2014} for density and temperature profiles in these clusters). Given typical plasma $\beta\sim 10^{2}$ in the ICM and sound wave $\lambda \sim$1 to 10 kpc 
(e.g., \citet{Fabian2006}), both the collisional and whistler regimes are expected to play a role in sound wave dissipation and propagation. 
Furthermore, the suppression of conduction expected in the whistler-dominated scattering regime may offer a natural explanation for the observed large propagation lengths of sound waves in the ICM.
Interestingly, \citet{Kunz20} demonstrated that suppression of collisionless Landau damping of ion acoustic waves is expected when the relative wave amplitudes exceed $2/\beta$. Such waves could then be self-sustaining and propagate over large distances in a manner resembling sound wave propagation in a weakly collisional ICM. 
%
%
%

An investigation of the consequences of whistler-mediated transport for the evolution of the AGN-induced waves represents an interesting future research direction, and we intend to report on the results of this investigation in future publications.

\subsection{Comparison with observations in the solar wind}
The solar wind is diffuse plasma flowing at high Mach number outward
in the heliosphere from the sun. {\it In situ} satellite observations have
produced enormous amounts of data on its properties and the Parker
Solar Probe mission will facilitate measurements as close as 10$R_\odot$
with $R_\odot$ the solar radius. The electron temperature falls slowly
with radial distance, from around 30~eV at 35$R_\odot$ to 10~eV at
120$R_\odot$ \citep{Moncuquet20}. The collisionality of the solar wind
depends on the plasma density and varies over a wide range, including a transition from collisional to collisionless behavior as the
collisional mean free path $L_c$ varies from smaller than to larger
than the scale length of the electron temperature gradient
$L$. Measurements from the large Wind spacecraft dataset revealed that
the electron heat flux rolled over to a value below the collisionless
heat flux $Q_0$ as $L_c/L$ increases and the ambient plasma becomes more collisionless \citep{Bale13}. While
$\beta_e$ of the solar wind is nominally of order unity around 1AU, this value is actually
highly variable. The Wind dataset revealed more than 12k measurements
of $\beta_e$ in the range from 5-100. An important conclusion from
this dataset in light of the threshold for whistler growth in
Eq.~(\ref{whistleronset}) is that value of $L_c/L$ above which the
electron heat flux rolled over to a constant value decreased with
higher $\beta_e$. An important development was the confirmation from
the Wind \citep{Tong18} and ARTEMIS \citep{Tong19} datasets that the
ratio of $Q/Q_0$ scales like $\beta_e^{-1}$ at high $\beta_e$,
consistent with whistler-limited heat flux. Further, the measured
amplitude of whistler waves increased both with heat flux and with
$\beta_e$ \citep{Tong19}, consistent with the heat flux as the
whistler drive mechanism.

While the general features of the whistler wave activity and the
associated heat flux measurements support the idea that heat flux
driven whistlers play a role in limiting electron heat flux in the
solar wind, significant uncertainties remain. The ARTEMIS observations
have been interpreted as ``quasi-parallel'' whistler waves
\citep{Tong19}. The PIC simulations as well as analytic analysis,
however, have established that parallel whistlers are not capable of
significantly limiting electron heat flux since the electrons carrying
the dominant heat flux do not resonate with parallel whistlers
\citep{Roberg-Clark16,Komarov18}. On the other hand, the magnetic
field measurements are limited to the spin plane of the spacecraft so
no direct measurements of the direction of the wavevectors of the
whistlers was possible. The amplitude of measured whistlers in the
solar wind was small, around $2\%$ of the ambient magnetic field,
leading to concern that the whistler waves were too small in amplitude
to limit electron thermal transport. On the other hand, the saturated
level of fluctuations given below Eq.~(\ref{nu_whistler}),
$\varepsilon_w\sim 10(\rho_e/L)nT$, is very small for the realistic
values of $\rho_e/L$ of the solar wind. For $T_e\sim 10~\mathrm{eV}$ and
$B_0\sim 10^{-4}~\mathrm{G}$, $\rho_e\sim 1~\mathrm{km}$. Taking a temperature scale
length of around $100R_\odot\sim 10^5~\mathrm{km}$ and $\beta_e\sim 10$, the
predicted whistler fluctuation level $\delta B/B_0\sim 1\%$, which is
in the range of the observations. The scaling $Q/Q_0\sim \beta_e^{-1}$
at high $\beta_e$ is now firmly established
\citep{Tong18,Tong19}. There has been no mechanism other than heat
flux limited whistlers proposed to explain this scaling.

The more recent measurements from the PSP mission have further established the role of whistlers in limiting the heat flux in the solar wind and specifically scattering the field aligned electron strahl, which has energies in the range of 100eV to 1keV, into the more isotropic halo distribution \citep{Agapitov20,Cattell21,Cattell21a}. The presence of large amplitude whistlers was correlated with local regions of increased plasma $\beta$ \citep{Agapitov20,Cattell21}. The presence of whistlers also correlated well with thresholds of fan instability, which is an oblique whistler driven by the $n=-1$ resonance \citep{Vasko19,Cattell21}. Further, the pitch angle width of the measured strahl electrons as also linked to the presence or absence of large amplitude whistlers, with the pitch angle width increasing with the strength of whistler wave activity, establishing that strahl scattering is caused by resonant interactions with whistlers. 

\begin{deluxetable*}{@{\extracolsep{4pt}}@{\kern\tabcolsep}cc>{\hspace{-12pt}}cccc}
\tabletypesize{\footnotesize}
\tablenum{1}
\tablecaption{\label{alpha_params}}
\tablehead{
\colhead{\textsc{Term}} & \colhead{\textsc{Definition}} & \multicolumn{2}c{ \textsc{When
  $\nu_{ei}^e \gg \nu_{we}$}} & \multicolumn{2}c{ \textsc{When $\nu_{ei}^e=0$}}\\
\cline{3-4}
\cline{5-6}
\colhead{} & \colhead{} &\colhead{General} & \colhead{$\gamma = \frac{4}{3}$} & \colhead{General} & \colhead{$\gamma = \frac{4}{3}$}
}
\decimalcolnumbers
\startdata
$\alpha_1$ &
$\frac{2}{3}\bigl<\frac{mv^2}{2T_0}\frac{\nu_{e}}{\nu}\bigr>$ &
$\frac{4}{3\sqrt{\pi}}\Gamma(4)$ & $4.51$&
$\frac{4}{3\sqrt{\pi}}\Gamma\bigl(\frac{5-\gamma}{2}\bigr)$ & $0.708$\\ 
$\alpha_2$ & $\frac{2}{3}\bigl<\frac{mv^2}{2T_0}\bigl(
\frac{mv^2}{2T_0} -\frac{5}{2} \bigr)\frac{\nu_{e}}{\nu}\bigr>$
& $\frac{2}{\sqrt{\pi}}\Gamma(4)$ & $6.77$ &
$-\frac{2}{3\sqrt{\pi}}\gamma\,\Gamma\bigl(\frac{5-\gamma}{2}\bigr)$ &$-0.472$\\ 
$\alpha_3$ & $\frac{2}{3}\bigl<\frac{mv^2}{2T_0}\frac{\nu_e}{\nu}\frac{\nu_w}{\nu_{we}}\bigr>$ 
& $\frac{4}{3\sqrt{\pi}}\Gamma\bigl(\frac{8+\gamma}{2}\bigr)$ & $11.1$ & $1$ & $1$\\ 
$\alpha_4$ &
$\frac{2}{3}\bigl<\bigl(\frac{mv^2}{2T_0}\bigr)^{\!\scriptscriptstyle{2}}\,\frac{\nu_{e}}{\nu}\bigr>$
& $\frac{4}{3\sqrt{\pi}}\Gamma(5)$ & $18.1$&
  $\frac{4}{3\sqrt{\pi}}\Gamma\bigl(\frac{7-\gamma}{2}\bigr)$ & $1.30$\\ 
$\alpha_5$ &
  $\frac{2}{3}\bigl<\bigl(\frac{mv^2}{2T_0}\bigr)^{\!\scriptscriptstyle{2}}\,\bigl(\frac{mv^2}{2T_0}
  -\frac{5}{2}\bigr)\frac{\nu_{e}}{\nu}\bigr>$ &
  $\frac{10}{3\sqrt{\pi}}\Gamma(5)$ & $45.1$&
    $\frac{4}{3\sqrt{\pi}}(1-\frac{\gamma}{2})\Gamma\bigl(\frac{7-\gamma}{2}\bigr)$ & $0.432$\\ 
$\alpha_6$ &
    $\frac{2}{3}\bigl<\bigl(\frac{mv^2}{2T_0}\bigr)^{\!\scriptscriptstyle{2}}\,\frac{\nu_e}{\nu}\frac{\nu_w}{\nu_{we}}\bigr>$ &
    $\frac{4}{3\sqrt{\pi}}\Gamma\bigl(\frac{10+\gamma}{2}\bigr)$ & $51.6$ &
    $\frac{5}{2}$ & $2.5$\\ 
$\alpha_7$ &
    $\frac{2}{3}\bigl<\frac{mv^2}{2T_0}\frac{\nu_e}{\nu}\frac{\nu_w^2}{\nu_{we}^2}\bigr>$
    & $\frac{4}{3\sqrt{\pi}}\Gamma(4+\gamma)$ & $30.2$&
    $\frac{4}{3\sqrt{\pi}}\Gamma\bigl(\frac{5+\gamma}{2}\bigr)$ & $1.76$\\ 
$\alpha_8$ & $\frac{2}{3}\bigl<\frac{mv^2}{2T_0}\bigl(
      \frac{mv^2}{2T_0}-\frac{5}{2}\bigr)\frac{\nu_e}{\nu}\frac{\nu_w}{\nu_{we}}\bigr>$
      & 
      $\frac{2}{3\sqrt{\pi}}\bigl(3+\gamma\bigr)\Gamma\bigl(\frac{8+\gamma}{2}\bigr)$
      & $24.0$&$0$ & $0$\\ 
$\alpha_9$ &
    $\frac{2}{3}\bigl<\frac{mv^2}{2T_0}\frac{\nu_{w}}{\nu_{we}}\bigr>$
    & $\frac{4}{3\sqrt{\pi}}\Gamma\bigl(\frac{5+\gamma}{2}\bigr)$ & $1.76$&
    $\frac{4}{3\sqrt{\pi}}\Gamma\bigl(\frac{5+\gamma}{2}\bigr)$ & $1.76$\\ 
$\alpha_{10}$ &
      $\alpha_8-\frac{\alpha_3\alpha_2}{\alpha_1}=\alpha_6-\frac{\alpha_3\alpha_4}{\alpha_1}$ &
      $\frac{2}{3\sqrt{\pi}}\gamma\,\Gamma\bigl(\frac{8+\gamma}{2}\bigr)$
      & $7.38$ &$\frac{\gamma}{2}$ & $0.667$\\ 
$\alpha_{11}$ &
      $\alpha_9 + \frac{\nu_{we}}{\nu_e}\bigl(\frac{\alpha_3^2}{\alpha_1}-\alpha_7\bigr)$ &
      $\frac{4}{3\sqrt{\pi}}\Gamma\bigl(\frac{5+\gamma}{2}\bigr)$ & $1.76$ &
      $\Gamma\bigl(\frac{5}{2}\bigr)/\Gamma\bigl(\frac{5-\gamma}{2}\bigr)$ & $1.41$\\ 
$\alpha_{12}$ &
      $\alpha_5-\frac{\alpha_2}{\alpha_1}$ &
      $\frac{10}{3\sqrt{\pi}}\Gamma(5)-\frac{3}{2}$ & $43.6$&
      $\bigl(1-\frac{\gamma}{2}\bigr)\Gamma\bigl(\frac{7-\gamma}{2}\bigr)/
      \Gamma\bigl(\frac{5}{2}\bigr)+\frac{\gamma}{2}$ & $1.10$\\ 
$\alpha_{13}$ &
      $\alpha_7-\frac{\alpha_3^2}{\alpha_1}$ &
      $\frac{4}{3\sqrt{\pi}}\bigl(\Gamma(4+\gamma)-\Gamma^2\bigl(\frac{8+\gamma}{2}\bigr)/6\bigr) $ & $3.05$ &
      $\frac{4}{3\sqrt{\pi}}\Gamma\bigl(\frac{5+\gamma}{2}\bigr)-\frac{3\sqrt{\pi}}{4}\Gamma^{-1}\bigl(\frac{5-\gamma}{2}\bigr) $ & $0.351$\\  
\enddata
\tablecomments{A table of the dimensionless parameters
  defining the electron transport equations. The brackets denote an
  average over a 3D Maxwellian distribution with temperature
  $T_0$. The parameters are evaluated in the limits when classical
  scattering dominates whistler scattering ($\nu_{ei}^e\gg\nu_{we}$) where
  $\nu\propto v^{-3}$ and when whistler scattering dominates classical
  scattering ($\nu_{we}\gg\nu_{ei}^e$) where $\nu\propto v^\gamma$.  Numerical values are given for the limits when $\gamma = 4/3$.}
\end{deluxetable*}

\begin{acknowledgments}

The authors thank the KITP program {\em Multiscale Phenomena in Plasma Astrophysics} for an inspiring hospitality that fostered the ideas for this work. C.P.\ thanks Thomas Berlok for valuable comments on the manuscript and enlightening discussions. This research was supported in part by the National Science Foundation under Grant No.\ NSF PHY-1748958. J.F.D., M.S.\ and A.E.\  acknowledge support from NASA ATP Grant No.\ NNX17AG27G, NASA Grant No. 80NNSC19K0848 and NSF Grant No.\ PHY1805829. C.P.\ and T.T.\ acknowledge support by the European Research Council (ERC) under ERC-CoG grant CRAGSMAN-646955.  C.S.R.\ acknowledges support by the ERC under ERC-Adv grant DISKtoHALO-834203. M.R.\ acknowledges support from 
NSF grants AST 1715140 and AST 2009227.

\end{acknowledgments}

\appendix
\section{Global coupled MHD and electron transport equations}
\label{appendix}
The transport equations presented in Sec.~\ref{overview} provide a
suitable framework when combined with an MHD description to describe
the large-scale dynamics of the ICM system. However, for simplicity
the equations were discussed in the context of a unidirectional
temperature gradient with whistler waves propagating in a single
direction down the gradient. In a real system the temperature will
develop complex structure that will produce whistlers propagating in
both directions with respect to the magnetic field. Thus, a set of
equations that can be used to describe the full dynamics of a system
with an arbitrary temperature structure must include bi-directional
whistler waves and their interaction with the ambient temperature
gradient. We therefore generalize the equations presented in
Sec.~\ref{overview} to describe whistlers propagating in both
directions with respect to the local magnetic field in three-dimensional space. The
generalization is straightforward except for the addition of a cross
term arising from $C_1(f_1)$ in Eq.~(\ref{equation:f2}). The cross
term describes electron heating associated with the interaction of
counter-streaming whistler waves. This has been interpreted as second
order Fermi acceleration in the case of cosmic ray transport
\citep{Thomas19}. The generalized equations when combined with MHD are
suitable for describing the dynamics of a system with arbitrary
gradients and arbitrary classical collision rates. We use the Lorentz-Heaviside system of units throughout this Appendix and denote the dyadic product of any two vectors $\mathbf{P}$ and $\mathbf{Q}$ by $\mathbf{P} \mathbf{Q}$.

\subsection{Derivation of the wave energy equation}

For the sake of transparency of this derivation, we suppress the source terms in the whistler wave equation, namely the whistler wave growth and loss terms (due to drag and second-order Fermi acceleration). Moreover, in this derivation we only consider whistler waves that move in one direction and generalize the result at the end of this section to also account for backwards moving whistlers. 

The energy equation for electromagnetic fields is:
\begin{equation}
    \frac{\partial (\mathbf{E}^2 / 2 + \mathbf{B}^2 / 2)}{\partial t} + c \nabla \cdot (\mathbf{E} \mathbf{\times} \mathbf{B}) + \mathbf{J} \cdot \mathbf{E} = 0,
    \label{eq:em_energy_eq}
\end{equation}
where the electric field in Hall-MHD (which is the appropriate limit) is given by:
\begin{equation}
    \mathbf{E} + \frac{[\mathbf{U} - \mathbf{J}/(ne)] \times \mathbf{B}}{c} = 
    \mathbf{0}.
\end{equation}
The Hall term, which is proportional to $\mathbf{J}/(ne)$, is assumed to be zero in standard MHD. This term does not do any work on the electro-magnetic field as can be shown:
\begin{align}
    \mathbf{J} \cdot \mathbf{E} &=  -\mathbf{J} \cdot \frac{[\mathbf{U} - \mathbf{J}/(ne)] \times \mathbf{B}}{c} =
    -\mathbf{J} \cdot \frac{\mathbf{U} \times \mathbf{B}}{c} =
    \mathbf{U} \cdot \frac{\mathbf{J} \times \mathbf{B}}{c} \\ &=
    \mathbf{U} \cdot [(\nabla \times \mathbf{B}) \times \mathbf{B}]
    = -\mathbf{U} \cdot \left[ \nabla \cdot \left( \frac{\mathbf{B}^2}{2} \mat{1} - \mathbf{B} \mathbf{B} \right)\right],
\end{align}
where we use vector identities in the last step and assumed a divergence-free magnetic field.

We are interested in the energy transport of small-amplitude whistler waves. To this end, we perturb the energy equation in Eq.~\eqref{eq:em_energy_eq} assuming small deviations of the electric and magnetic fields from its mean values. Carrying out a perturbation analysis and neglecting the energy density of the electric field in the energy equation yields:
\begin{equation}
    \frac{\partial \delta \mathbf{B}^2 / 2}{\partial t} + c \nabla \cdot (\delta \mathbf{E} \mathbf{\times} \delta\mathbf{B}) + \delta \mathbf{J} \cdot \delta \mathbf{E} = 0,
\end{equation}
and the perturbed electric field for whistler waves is given by:
\begin{equation}
    \delta \mathbf{E} + \frac{(\mathbf{U} + \mathbf{V}_w) \times \delta\mathbf{B}}{c} = \mathbf{0}, \label{eq:electric_field}
\end{equation}
where $\mathbf{V}_w$ is the whistler wave velocity.
Inserting the perturbed electric field into the divergence terms and expanding the double-cross product yields:
\begin{equation}
    \frac{\partial \delta \mathbf{B}^2 / 2}{\partial t} - \nabla \cdot \left[(\mathbf{U} + \mathbf{V}_w) \cdot\delta\mathbf{B}\delta\mathbf{B} - \delta\mathbf{B}^2(\mathbf{U} + \mathbf{V}_w)\right] + \delta \mathbf{J} \cdot \delta \mathbf{E} = 0.\label{eq:wave_energy_1}
\end{equation}
The magnetic field tensor is given by:
\begin{align}
\delta\mathbf{B}\delta\mathbf{B} &= 
\left[\begin{matrix} \delta B_x \delta B_x & \delta B_y \delta B_x & 0 \\ \delta B_x \delta B_y & \delta B_y \delta B_y &  0 \\ 0 & 0 & 0 \end{matrix}\right]
\end{align}
where we assumed that the whistler waves are propagating quasi-parallel to the mean magnetic field and evaluated the tensor in a coordinate system where this mean magnetic field is locally aligned with the $z$ axis.
In general the whistler wave has some phase. As we have no a priori information about this phase, we take an average of the magnetic field tensor over all possible phases and assume an equal probability for all phases. This results in
\begin{align}
\left\langle \delta\mathbf{B}\delta\mathbf{B} \right\rangle &= \left[\begin{matrix} \delta \mathbf{B}^2 /2 & 0 & 0 \\ 0 & \delta \mathbf{B}^2 /2 & 0 \\ 0 & 0 & 0 \end{matrix}\right] 
= \frac{\delta \mathbf{B}^2}{2} (\mat{1} - \mathbf{b}\mathbf{b}),
\label{eq:magnetic_square_tensor}
\end{align}
where $\langle\rangle$ denotes the ensemble average, $\delta B_x = |\delta \mathbf{B}| \cos(\phi)$, $\delta B_y = \pm |\delta \mathbf{B}| \sin(\phi)$, $\mathbf{b} = \mathbf{B} / |\mathbf{B}|$ is the unit vector along ${\bf B}$, and we applied the ergodic theorem so that we effectively replace ergodic averaging by phase averaging.\footnote{Note that $1/(2\pi) \int_{-\pi}^{\pi} d\phi \sin^{2}(\phi) = 1/(2\pi) \int_{-\pi}^{\pi} d\phi \cos^{2}(\phi) = 1/2$.} Taking the ensemble average of Eq.~\eqref{eq:wave_energy_1} and inserting the result of Eq.~\eqref{eq:magnetic_square_tensor} yields:
\begin{equation}
    \frac{\partial \delta \mathbf{B}^2 / 2}{\partial t} - \nabla \cdot \left[(\mat{1} - \mathbf{b}\mathbf{b})\cdot (\mathbf{U} + \mathbf{V}_w) \frac{\delta \mathbf{B}^2}{2}  - \delta\mathbf{B}^2(\mathbf{U} + \mathbf{V}_w)\right] + \left\langle \delta \mathbf{J} \cdot \delta \mathbf{E} \right\rangle= 0
\end{equation}
After noting that $(\mat{1} - \mathbf{b} \mathbf{b})\cdot \mathbf{V}_w = 0$ we get:
\begin{equation}
    \frac{\partial \delta \mathbf{B}^2 / 2}{\partial t} + \nabla \cdot \left[\delta\mathbf{B}^2(\mathbf{U} + \mathbf{V}_w) - (\mat{1} - \mathbf{b}\mathbf{b}) \cdot \mathbf{U} \frac{\delta \mathbf{B}^2}{2}\right] + \left\langle \delta \mathbf{J} \cdot \delta \mathbf{E} \right\rangle= 0.
    \label{eq:wave_energy_2}
\end{equation}
The work done on the perturbed electro-magetic field is:
\begin{align}
\delta \mathbf{J} \cdot \delta \mathbf{E}
&= -\mathbf{U} \cdot \left[ \nabla \cdot \left( \frac{\delta\mathbf{B}^2}{2} \mat{1} - \delta \mathbf{B} \delta \mathbf{B}\right) \right],
\end{align} 
or, after taking the ensemble average (using Eq.~\ref{eq:magnetic_square_tensor}),
\begin{align}
\left\langle \delta \mathbf{J} \cdot \delta \mathbf{E} \right\rangle
&= -\mathbf{U} \cdot \left[ \nabla \cdot \left( \frac{\delta\mathbf{B}^2}{2} \mathbf{b} \mathbf{b}\right) \right].
\end{align}
Inserting this expression into Eq.~\eqref{eq:wave_energy_2} yields:
\begin{equation}
    \frac{\partial \delta \mathbf{B}^2 / 2}{\partial t} + \nabla \cdot \left[\frac{\delta \mathbf{B}^2}{2} \mathbf{U}  + \frac{\delta\mathbf{B}^2}{2} \mathbf{b}\mathbf{b} \cdot \mathbf{U}  + \delta \mathbf{B}^2 \mathbf{V}_w\right] = \mathbf{U}  \cdot \left[ \nabla \cdot \left( \frac{1}{2}\delta\mathbf{B}^2 \mathbf{b} \mathbf{b}\right) \right].
    \label{eq:wave_energy_3}
\end{equation}
Defining the energy density and pressure tensor of whistler waves, respectively,
\begin{align}
    \varepsilon_w &= \frac{\delta \mathbf{B}^2}{2} ,~\mbox{and}\\
    \mat{P}_w &= \frac{\delta \mathbf{B}^2}{2} \mathbf{b}\mathbf{b}
\end{align}
simplifies the notation of Eq.~\eqref{eq:wave_energy_3}:
\begin{equation}
    \frac{\partial\varepsilon_w}{\partial t} + \nabla \cdot \left[  (\varepsilon_w \mat{1}+ \mat{P}_w) \cdot \mathbf{U}  + 2 \varepsilon_w \mathbf{V}_w\right] = \mathbf{U} \cdot \left[ \nabla \cdot \mat{P}_w \right].
\end{equation}
In order to generalize this result to forward and backward propagating whistlers, we define the corresponding energy densities, $\varepsilon_w^\pm$, and wave pressure tensor, $\mat{P}_w^\pm$, add the whistler source terms to the right-hand side, and obtain\footnote{Using the identity
$\nabla \cdot \left[\mat{P}_w^\pm
\cdot \mathbf{U} \right] - \mathbf{U} \cdot \left[\nabla \cdot
\mat{P}_w^\pm \right]
= \mat{P}_w^\pm \mathbf{:} \nabla \mathbf{U}$,
this equation can also be written
\begin{equation}
    \frac{\partial\varepsilon_w^\pm}{\partial t} + \nabla \cdot \left[ \varepsilon_w^\pm \mathbf{U}  \pm 2 \varepsilon_w^\pm \mathbf{V}_w\right] = -\mat{P}_w^\pm \mathbf{:} \nabla \mathbf{U} + G_w^\pm - H_w^\pm.
\end{equation}
}:
\begin{equation}
    \frac{\partial\varepsilon_w^\pm}{\partial t} + \nabla \cdot \left[  (\varepsilon_w^\pm \mat{1}+ \mat{P}_w^\pm) \cdot \mathbf{U}  \pm 2 \varepsilon_w^\pm \mathbf{V}_w\right] = \mathbf{U} \cdot \left[ \nabla \cdot \mat{P}_w^\pm \right] + G_w^\pm - H_w^\pm.
    \label{eps_w}
\end{equation}
The superscript $\pm$ denotes waves propagating down and up the temperature gradient, respectively. The whistler energy density and pressure tensor are given by
\begin{align}
    \varepsilon_w^\pm &= \frac{\delta \mathbf{B}_\pm^2}{2},~\mbox{and} \\
    \mat{P}_w^\pm &= \frac{\delta \mathbf{B}_\pm^2}{2} \mathbf{b}\mathbf{b}.
\end{align}
The whistler growth term, $G_w^\pm$, and the whistler loss term, $H_w^\pm$, are given by
\begin{align}
G_w^\pm &= \pm V_{\mathrm{st}}^\pm n k_\mathrm{B} \mathbf{b} \cdot \mathbf{\nabla} T,\\
H_w^\pm &=   m_e n V_w^2
    \left(\alpha_{11}\nu_{we}^\pm + \alpha_{13} \frac{\nu_{we}^+
    \nu_{we}^-}{\nu_e}\right),
\end{align}
where $k_\mathrm{B}$ is Boltzmann's constant. The first term in the whistler loss term $H_w^\pm$ is associated with the wave drag on the electrons that causes them to be heated and includes contributions from whistlers propagating in both directions. The term proportional to the product of the scattering rates of bi-direction whistlers corresponds to second order Fermi acceleration. The generalized electron streaming velocities $V_{\mathrm{st}}^\pm$ with the forward and backward propagating whistler waves are defined as
\begin{equation}
    V_{\mathrm{st}}^\pm = V_w \alpha_{10}
    \frac{\nu_{we}^\pm}{\nu_e} \,,
\end{equation}
where $V_w=\mathbf{V}_w\cdot\mathbf{b}=v_{te}/\beta_e$ is the non-directional whistler phase speed.

\subsection{Momentum conservation and kinetic energy equation}

Analogously, it is possible to perturb the momentum equation to account for the forces exerted by the whistler waves. We start with the momentum equation for the composite ion-electron fluid, which is given by
\begin{align}
    \rho \frac{d \mathbf{U}}{d t} + \nabla P_\mathrm{th} = \frac{\mathbf{J} \times \mathbf{B}}{c} 
    = (\nabla \times \mathbf{B}) \times \mathbf{B} 
    = - \nabla \cdot \left( \frac{\mathbf{B}^2}{2} \mat{1} - \mathbf{B} \mathbf{B} \right),
\end{align}
where where $d/dt=\partial /\partial t+{\bf U}\cdot {\bf\nabla}$ is the Lagrangian derivative, $P_\mathrm{th} = P_e + P_i$ is the total (ion plus electron) pressure, and we used $\mathbf{J} = c \nabla \times \mathbf{B}$ (neglecting the displacement current in the Hall-MHD approximation). Introducing the perturbations in the magnetic field while neglecting linear contributions due to ensemble averaging yields:
\begin{align}
    \rho \frac{d \mathbf{U}}{d t} + \nabla\cdot \left( P_\mathrm{th} \mat{1} + \frac{\mathbf{B}^2}{2} \mat{1} - \mathbf{B} \mathbf{B} + \frac{\delta \mathbf{B}^2}{2} \mat{1} - \langle \delta \mathbf{B} \delta\mathbf{B} \rangle \right) &= \mathbf{0},
\end{align}
which can be simplified using Eq.~\eqref{eq:magnetic_square_tensor} to yield:
\begin{align}
    \rho \frac{d \mathbf{U}}{d t} + \nabla\cdot \left( P_\mathrm{th} \mat{1} + \frac{\mathbf{B}^2}{2} \mat{1} - \mathbf{B} \mathbf{B} + \frac{\delta \mathbf{B}^2}{2} \mathbf{b}\mathbf{b} \right) &= \mathbf{0}.
\end{align}
Using the definitions for the wave energy and wave pressure while accounting for the two wave types and inserting the continuity equation yields the final conservative form of the momentum equation in the presence of whistler waves:
\begin{equation}
    \frac{\partial\rho \mathbf{U}}{\partial t} + \mathbf{\nabla} \cdot \left(  \rho \mathbf{U} \mathbf{U} + P_\mathrm{th} \mat{1} + \frac{\mathbf{B}^2}{2}\mat{1} - \mathbf{B} \mathbf{B} +\mat{P}_w^+ +\mat{P}_w^- \right) = \mathbf{0}.
\end{equation}
Multiplying this equation by $\mathbf{U}$ and using the continuity equation yields the kinetic energy equation: 
\begin{equation}
    \frac{\partial \varepsilon_\mathrm{kin}}{\partial t} + \mathbf{\nabla} \cdot \left(\varepsilon_\mathrm{kin} \mathbf{U} \right) = - \mathbf{U} \cdot \left\{\mathbf{\nabla} \cdot \left[ \left(P_\mathrm{th} + \frac{\mathbf{B}^2}{2}\right)\mat{1} -\mathbf{B} \mathbf{B} + \mat{P}_w^++ \mat{P}_w^- \right] \right\}.
    \label{eps_kin}
\end{equation}
where $\varepsilon_\mathrm{kin}=\rho \mathbf{U}^2 / 2$.

\subsection{Magnetic energy equation}

The magnetic induction law is given by,
\begin{equation}
    \frac{\partial \mathbf{B}}{\partial t} + \mathbf{\nabla} \cdot (\mathbf{B} \mathbf{U} -     \mathbf{U} \mathbf{B}) = \mathbf{0}.
\end{equation}
Multiplying this equation by $\mathbf{B}$ yields the equation for the magnetic energy:
\begin{equation}
    \frac{\partial \varepsilon_B}{\partial t} + \mathbf{\nabla} \cdot \left[\mathbf{U}\cdot\left(\varepsilon_B+\frac{\mathbf{B}^2}{2}\right)- (\mathbf{U} \cdot \mathbf{B})\mathbf{B}\right] =  \mathbf{U} \cdot \left\{\mathbf{\nabla} \cdot \left[ \frac{\mathbf{B}^2}{2}\mat{1} -\mathbf{B} \mathbf{B} \right] \right\},
    \label{eps_B}
\end{equation}
where $\varepsilon_B=\mathbf{B}^2/2$.

\subsection{Ion and electron energy equations}
The ion and electron energy equations are
\begin{align}
    \frac{\partial\varepsilon_i}{\partial t}
    +
    \mathbf{\nabla} \cdot \left[\mathbf{U}(\varepsilon_i+ P_i)\right]
    &=
    \mathbf{U} \cdot \mathbf{\nabla} P_i  
    - \frac{\varepsilon_i - \varepsilon_e}{\tau_{\mathrm{eq}}} ,    \label{eps_i}\\
    \frac{\partial\varepsilon_e}{\partial t}
    +
    \mathbf{\nabla} \cdot \left[\mathbf{U}(\varepsilon_e+P_e)
    + \mathbf{Q}_{e,\mathrm{st}}
    + \mathbf{Q}_{e,\mathrm{dif}}
    \right]
    &=
    \mathbf{U} \cdot \mathbf{\nabla} P_e  
    - \frac{\varepsilon_e - \varepsilon_i}{\tau_{\mathrm{eq}}} 
    + \sum_\pm ( H_w^\pm - G_w^\pm ).
    \label{eps_e}
\end{align}
where $\varepsilon_i=3/2 n k_\mathrm{B} T_i$ and $\varepsilon_e=3/2 n k_\mathrm{B} T_e$ are the energy densities of ions and electrons, respectively, and the electron streaming and diffusion fluxes are given by
\begin{align}
    \mathbf{Q}_{e,\mathrm{st}} &= (V_\mathrm{st}^+ - V_\mathrm{st}^-) P_e\mathbf{b},\\
    \mathbf{Q}_{e,\mathrm{dif}} &= -\kappa_e\mathbf{b}\mathbf{b}\cdot \mathbf{\nabla} T_e.
\end{align}
While the ion energy equation is standard, the equation for the electron energy density is the same as given in Eq.~(\ref{energy_electrons}) but generalized for the general fluid velocity {\bf U} and the whistler streaming velocities $V_{\mathrm{st}}^\pm$. The energy equilibration time between electrons and ions is denoted by $\tau_{\mathrm{eq}}$, which is typically longer by $\sqrt{m_i/m_e}$ than the classical electron-ion scattering time. The whistler waves can also transfer energy between electrons and ions but the detailed scaling behavior for this transfer has not been established \citep{Roberg-Clark18a}. Viscous terms could also be included in the momentum and ion pressure equations. The parallel conductivity $\kappa_e$ is
\begin{equation}
    \kappa_e=\alpha_{12}\frac{nk_\mathrm{B}T_e}{m_e\nu_e}.
      \label{conductivityA}
  \end{equation}
The total scattering rate $\nu_e=\nu_{ei}^e+\nu_{we}^++\nu_{we}^-$ is the sum of $\nu_{ei}^e$, the classical electron-ion scattering rate, and the scattering rates $\nu_{we}^\pm$ from the forward and backward propagating whistlers, all evaluated at the electron thermal velocity $v_{te}$, 
  \begin{equation}
    \nu_{ei}^e=\frac{4\pi e^4n\Lambda}{m_e^2v_{te}^3}, \;\; \nu_{we}^\pm=0.1\Omega_{e}\frac{\varepsilon_w^\pm}{\varepsilon_B}, 
      \label{nuA}
  \end{equation}
where $\varepsilon_w^\pm$ and $\varepsilon_B =B^2/8\pi$ are the whistler and magnetic energy densities and $\Lambda$ is the Coulomb logarithm.

The total electron heat flux $\mathbf{Q}_{e} = \mathbf{U}(\varepsilon_e+P_e) + \mathbf{Q}_{e,\mathrm{st}} + \mathbf{Q}_{e,\mathrm{dif}}$ is composed of the flux due to advection of electron enthalpy, the effective streaming flux due to electron-whistler wave scattering, as well as the diffusive flux. In Eq.~(\ref{eps_e}), the forward and backward propagating whistlers try to carry the electron energy in opposite directions along ${\bf B}$. The wave with the dominant scattering rate wins out and as a result the direction of advection can change sign with the ambient parallel temperature gradient. 

\subsection{Full set of equations of whistler MHD}
\label{sec:W-MHD}

For completeness, here we summarize the complete set of equations for whistler MHD that constitute a complete description of plasma dynamics in the high $\beta$ ICM (to which a description of anisotropic viscosity can be added):
\begin{align}
\frac{\partial \rho}{\partial t} + \mathbf{\nabla} \cdot (\rho \mathbf{U}) &= 0, \label{eq:continuityA}\\
\frac{\partial \rho \mathbf{U}}{\partial t} + \mathbf{\nabla} \cdot \left[\rho \mathbf{U} \mathbf{U} + \left(P_\mathrm{th} + \frac{\mathbf{B}^2}{2}\right)\mat{1} -     \mathbf{B} \mathbf{B} +\mat{P}_w^+ +\mat{P}_w^-\right] &= \mathbf{0},     \label{eq:final_gas_euler}\\
\frac{\partial \mathbf{B}}{\partial t} + \mathbf{\nabla} \cdot (\mathbf{B} \mathbf{U} -     \mathbf{U} \mathbf{B}) &= \mathbf{0}, \label{eq:induction_law} \\
\frac{\partial\varepsilon_i}{\partial t}
    + \mathbf{\nabla} \cdot \left[\mathbf{U}(\varepsilon_i+P_i)\right]
&=  \mathbf{U} \cdot \mathbf{\nabla} P_i 
    - \frac{\varepsilon_i - \varepsilon_e}{\tau_{\mathrm{eq}}}, \label{eq:eps_i2}\\
\frac{\partial\varepsilon_e}{\partial t}
    +\mathbf{\nabla} \cdot \left[\mathbf{U}(\varepsilon_e+P_e)
    + \mathbf{Q}_{e,\mathrm{st}}
    + \mathbf{Q}_{e,\mathrm{dif}}
    \right]
    &= \mathbf{U} \cdot \mathbf{\nabla} P_e  
    - \frac{\varepsilon_e - \varepsilon_i}{\tau_{\mathrm{eq}}}
    + \sum_\pm ( H_w^\pm - G_w^\pm ),
    \label{eq:eps_e2}\\\
\frac{\partial\varepsilon_w^\pm}{\partial t} 
    + \nabla \cdot \left[  (\varepsilon_w^\pm \mat{1}+ \mat{P}_w^\pm) \cdot \mathbf{U}  \pm 2 \varepsilon_w^\pm \mathbf{V}_w\right] 
    &=  \mathbf{U} \cdot \left[ \nabla \cdot \mat{P}_w^\pm \right] + G_w^\pm - H_w^\pm.
\end{align}
where $P_\mathrm{th}=P_i+P_e$ and the equations of state are given by
\begin{align}
    P_i &= n k_\mathrm{B} T_i =  (\Gamma - 1) \varepsilon_i,\\
    P_e &= n k_\mathrm{B} T_e = (\Gamma - 1) \varepsilon_e,
\end{align}
where $\Gamma=5/3$. The whistler growth term, $G_w^\pm$, and the whistler loss terms, $H_w^\pm$, (due to drag and second-order Fermi acceleration) are given by
\begin{align}
G_w^\pm &= \pm V_{\mathrm{st}}^\pm n k_\mathrm{B} \mathbf{b} \cdot \mathbf{\nabla} T_e,\\
H_w^\pm &=   m_e n V_w^2
    \left(\alpha_{11}\nu_{we}^\pm + \alpha_{13} \frac{\nu_{we}^+
    \nu_{we}^-}{\nu_e}\right).
    \label{eq:H_w}
\end{align}
The whistler energy density and pressure tensor are given by
\begin{align}
    \varepsilon_w^\pm &= \frac{\delta \mathbf{B}_\pm^2}{2}, \\
    \mat{P}_w^\pm &= \frac{\delta \mathbf{B}_\pm^2}{2} \mathbf{b}\mathbf{b}.
\end{align}

The parameters $\alpha_{i}$ with various subscripts in Eqs.~(\ref{eq:eps_e2})-(\ref{eq:H_w}) are given in Table \ref{alpha_params}. As shown in the Table these parameters have simple analytic forms in the limit of large or small classical collisions but no simple analytic form for arbitrary $\nu_{we}/\nu_{ei}^e$. The parameter $\gamma$ in the Table controls the dependence of $\nu_{w}$ on velocity, $\nu_{w}=\nu_{we}(v/v_{te})^\gamma$, where, as discussed in Sec.~\ref{derivation}, our best estimate is that $\gamma=4/3$. The values of the parameters $\alpha_i$ in the two collisionality limits have been explicitly evaluated for $\gamma=4/3$. A connection formula for the two collisionality limits could be constructed in a numerical implementation of the transport equations.  Further, as mentioned previously the thermal conduction in Eq.~(\ref{conductivityA}) does not reduce to the Spitzer value because we have discarded electron-electron collisions. The correct Spitzer value is obtained by setting $\alpha_{12}$ to 4.25 rather than the value given in Table \ref{alpha_params}.

Finally, we note that in a large-scale system in which the transport time scale $L/V_w$ is long compared the electron-ion energy exchange time $\tau_{\mathrm{eq}}$, Eqs.~\eqref{eq:eps_i2} and \eqref{eq:eps_i2} can be combined into a single energy equation for electrons and ions.

\subsection{Energy conservation}

Adding up Eqs.~\eqref{eps_w}, \eqref{eps_kin}, \eqref{eps_B}, \eqref{eps_i}, and \eqref{eps_e}, we obtain total energy conservation,
\begin{align}
    \frac{\partial\varepsilon_\mathrm{tot}}{\partial t}
    +
    \mathbf{\nabla} \cdot \left[\mathbf{U}\cdot(\varepsilon_\mathrm{tot}\mat{1}+\mat{P}_\mathrm{tot})
    -(\mathbf{U}\cdot\mathbf{B})\mathbf{B}
    + 2\mathbf{V}_w (\varepsilon_w^+-\varepsilon_w^-) 
    + \mathbf{Q}_{e,\mathrm{st}}
    + \mathbf{Q}_{e,\mathrm{dif}}
    \right]
    &=0
    \label{eps_tot}
\end{align}
where 
\begin{align}
\varepsilon_\mathrm{tot}&=\varepsilon_i+\varepsilon_e+\varepsilon_\mathrm{kin}+\varepsilon_B+\varepsilon_w^++\varepsilon_w^-,~\mbox{and}\\ \mat{P}_\mathrm{tot}&=\left(P_i+P_e+\frac{\mathbf{B}^2}{2}\right)\mat{1}
    +\mat{P}_w^++\mat{P}_w^-
    \label{def}
\end{align}
are the total energy density and the pressure tensor, respectively.

\newpage

\end{document}